\documentclass[twocolumn,preprintnumbers,superscriptaddress,nofootinbib,aps,prd,floatfix]{revtex4-1}
\pdfoutput=1 
\usepackage{multirow}
\usepackage[utf8]{inputenc}
\usepackage{graphicx}
\usepackage{amsmath,mathrsfs,amscd,graphicx}
\usepackage{color}
\usepackage{rotating}
\usepackage{slashed}
\usepackage{subfig}
\usepackage{textcomp}
\usepackage{array}
\usepackage{float}
\usepackage[hyphens,spaces,obeyspaces]{xurl}
\usepackage{textcomp}
\usepackage{amsfonts, latexsym, epsfig}
\usepackage{bm}
\usepackage{times}
\usepackage{epsfig}
\usepackage{amssymb}
\usepackage{tikz}
\usepackage{cancel}
\usepackage{hyperref}
\usepackage{verbatim}
\usepackage[normalem]{ulem}
\usepackage{diagbox}
\usepackage{xspace}

\def\beq{\begin{equation}}
\def\eeq{\end{equation}}
\def\bea{\begin{eqnarray}}
\def\eea{\end{eqnarray}}
\def\MCFM{{\tt MCFM}\xspace}

\newcommand{\CP}{\text{CP}\xspace} 
\newcommand{\PH}{H}
\newcommand{\ttH}{t\bar{t}\PH}

\newcommand{\tqH}{t{q}\PH}
\newcommand{\tWH}{t{W}\PH}
\newcommand{\ttb}{t\bar{t}}

\newcommand{\ifb}{\ensuremath{\mathrm{fb^{-1}}}}                                
\newcommand{\Mtt}{M_{t\bar{t}}}
\newcommand{\ytt}{\Delta y_{t\bar{t}}}
\newcommand{\pt}{p_{\mathrm{T}}}
\newcommand{\ptmiss}{p_{\mathrm{T}}^{\mathrm{miss}}}
\newcommand{\ptvecmiss}{\vec{p}_{\mathrm{T}}^{\mathrm{miss}}}

\begin{document}
\title{Probing the \CP structure of the top quark Yukawa coupling: Loop sensitivity vs. on-shell sensitivity}
\author{Till Martini}
\email{till.martini@physik.hu-berlin.de}
\affiliation{Humboldt-Universit\"at zu Berlin, Institut f\"ur Physik, Newtonstra{\ss}e 15, 12489 Berlin, Germany}
\author{Ren-Qi Pan}
\email{renqi.pan@cern.ch}
\affiliation{Zhejiang Institute of Modern Physics, Department of Physics, Zhejiang University, Hangzhou, Zhejiang 310027, China}
\author{Markus Schulze}
\email{markus.schulze@physik.hu-berlin.de}
\affiliation{Humboldt-Universit\"at zu Berlin, Institut f\"ur Physik, Newtonstra{\ss}e 15, 12489 Berlin, Germany}
\email{markus.schulze@physik.hu-berlin.de}
\author{Meng Xiao}
\email{meng.xiao@cern.ch}
\affiliation{Zhejiang Institute of Modern Physics, Department of Physics, Zhejiang University, Hangzhou, Zhejiang 310027, China}

\begin{abstract}
The question whether the Higgs boson is connected to additional \CP violation is one of the driving forces behind precision studies at the Large Hadron Collider.
In this work, we investigate the \CP structure of the top quark Yukawa interaction {\textemdash} one of the most prominent places for searching for {New Physics} {\textemdash} through Higgs boson loops in top quark pair production. 
We calculate the electroweak corrections including arbitrary \CP mixtures at next-to-leading-order in 
the Standard Model Effective Field Theory. 
This approach of probing Higgs boson degrees of freedom relies on the large $t\bar{t}$ cross section and the 
excellent perturbative control. 
In addition, we consider all direct probes with on-shell Higgs boson production in association with a single top quark or top quark pair. 
This allows us to contrast loop sensitivity versus on-shell sensitivity in these fundamentally different process dynamics. 
We find that loop sensitivity in  $t\bar{t}$ production and on-shell sensitivity in $\ttH$ and $tH$ provide complementary handles over a wide range of parameter space.
\end{abstract}

\maketitle

\section{Introduction}
\label{intro}
The discovery of the Higgs boson by the ATLAS and CMS Collaborations~\cite{Chatrchyan:2012ufa, Aad:2012tfa,Chatrchyan:2013lba} in 2012 has sparked an extensive research effort to precisely measure its properties and origin. In fact, it has become one of the main goals of collider phenomenology. 
Data shows that the discovered particle is consistent with the spin-zero Higgs boson of the Standard Model (SM)~\cite{Khachatryan:2014kca,Aad:2015zhl} {\textendash} within the current uncertainties. 
One defining property is the  \CP-even {\it Yukawa interaction} between the Higgs boson and the SM fermions, which is proportional to the fermion mass. 
\CP-violating contributions are in principle allowed by gauge as well as space-time symmetries, and they are theoretically compelling in the context of the baryon asymmetry in the universe.    
In fact, many generic extensions of the SM, such as two-Higgs-doublet models~\cite{Ginzburg:2004vp,Gunion:2005ja,Maniatis:2007vn,Branco:2011iw,Haber:2012np} or composite Higgs models~\cite{Kaplan:1983sm,Georgi:1984ef,Georgi:1984af,Dugan:1984hq,Contino:2003ve,Agashe:2004rs,Giudice:2007fh,Contino:2010rs,DeCurtis:2011yx,Redi:2011zi,Mrazek:2011iu,Redi:2012ha,Montull:2013mla,Panico:2015jxa,Erdmenger:2020lvq}, contain modifications of the simple \CP structure of the SM. 
It is therefore of highest importance to search for \CP-odd contributions that modify the SM interactions.
In this regard, the top quark Yukawa interaction stands out due to the large top quark mass, which implies a large coupling. 
Moreover, top quarks are copiously produced at the LHC, which makes them ideal for searching for {New Physics}. 
For example, the CMS and ATLAS experiments~\cite{Sirunyan:2020sum, Aad:2020ivc} recently  presented first measurements which resulted in the exclusion of 
a pure \CP-odd coupling at more than 3$\sigma$, still leaving large room for mixtures of \CP-even and \CP-odd components.

In this work, we propose a novel way to probe the \CP properties of the top quark Yukawa coupling:
We want to challenge the SM at loop level, where the Higgs boson only appears through off-shell degrees of freedom. 
Hence, we study top quark pair production accounting for electroweak corrections.  
The leading effects are the $\mathcal{O}(\alpha)$ weak corrections to the QCD-induced $pp\to t\bar t$ process, which we calculate for 
\CP-even {\it and} \CP-odd $Htt$ couplings allowing arbitrary mixtures. 
Our strategy is motivated by the abundant data of top quark pairs produced at the LHC and excellent perturbative control over the theoretical predictions.
In fact, the LHC will have produced almost 500 million top quarks by 2023, while already today, theoretical predictions reach an accuracy of only a few percent~\cite{Czakon:2013goa,Czakon:2015owf,Czakon:2016ckf,Gao:2017goi,Czakon:2017wor,Behring:2019iiv}. 
These prospects are only getting brighter with the high-luminosity runs commencing in 2027 and high ambitions of the theory community towards $\mathrm{N^3LO}$ calculations~\cite{Muselli:2015kba,Piclum:2018ndt,snowmass2021}.
Another promising feature is that this approach, using $t\bar{t}$, is free from {\it penalties} of Higgs boson branching fractions and ambiguities of a complicated final state, 
which is in contrast to on-shell Higgs production processes such as $pp\to t\bar{t}H$. 
Yet, it is at the same order in the perturbative counting. 

We note that this idea was actually presented for
the first time a long time ago in Ref.~\cite{Schmidt:1992et} proposing the difference in the transverse energy distribution of leptons and antileptons from $t\bar{t}$ events at hadron colliders as probe of \CP violation in the Higgs sector. A previous study~\cite{Kuhn:2013zoa} has also already touched upon this idea. 
The authors consider $t\bar{t}$ production at the LHC and allow rescaling the \CP-even Yukawa coupling in a pre-existing SM calculation. 
An actual measurement by CMS~\cite{Sirunyan:2019nlw} shows very promising sensitivity that warrants further investigation.
In contrast, our work requires calculating the respective Higgs boson loops from scratch due to the new \CP-odd components. 
We present a realistic phenomenological analysis for the most general \CP-even and \CP-odd coupling structure, for the first time, and estimate the sensitivity using  
the same observables as in the CMS analysis in Ref.~\cite{Sirunyan:2019nlw}.
In addition, we simulate the {competing} on-shell processes $pp\to \ttH$, $pp\to \tqH$ and $pp\to \tWH$ with the same \CP-even and \CP-odd top quark Yukawa couplings  
in order to have a fair comparison and to capture dominant backgrounds.
We partly resort to our previous work in Ref.~\cite{Gritsan:2016hjl} and extend it by the calculation of the $pp\to \tWH$ process,
which we discuss in more detail.
As a result, all on-shell processes are publicly available\footnote{\href{https://spin.pha.jhu.edu}{https://spin.pha.jhu.edu}.} in the {\tt JHUGen} Monte-Carlo generator~\cite{Gao:2010qx,Bolognesi:2012mm,Anderson:2013afp,Gritsan:2016hjl,Gritsan:2020pib}, 
which is heavily used in experimental analysis~\cite{Khachatryan:2014kca,Khachatryan:2016tnr,Sirunyan:2017tqd,Sirunyan:2019twz,Sirunyan:2019nbs,Sirunyan:2020sum}.
We also provide a publicly available\footnote{\href{https://github.com/TOPAZdevelop/MCFM-8.3\_EWSMEFT\_ADDON}{https://github.com/TOPAZdevelop/MCFM-8.3\_EWSMEFT\_ADDON}} extension of {\MCFM}~\cite{Campbell:2016dks,Campbell:2019dru}, which 
yields the loop correction to $t\bar{t}$ production. 
Finally, we note that our implementation also allows for the most general CP-even and CP-odd $HWW$ anomalous couplings in the tWH process. For the purpose of this work, however, we keep them at their SM value. 
Recently, the combination of all on-shell processes has also been studied in Ref.~\cite{Bahl:2020wee}. 
Relevant works on a subset of these processes can be found in 
Refs.~\cite{Ellis:2013yxa,Demartin:2014fia,Buckley:2015ctj,Demartin:2015uha,Demartin:2016axk,Kobakhidze:2016mfx,Azevedo:2017qiz,Barger:2018tqn,Kraus:2019myc,Faroughy:2019ird,Bortolato:2020zcg}.
Also low-energy measurements of the electric dipole moment yield complementary constraints on the \CP-odd components, which are 
remarkably strong~\cite{Brod:2013cka,Chien:2015xha,Cirigliano:2016nyn,Panico:2018hal,Fuchs:2020uoc}
and need to be considered in real data analyses.

\section{Loop sensitivity to the \CP structure of the top-Higgs coupling}
\label{ttbar}
\subsection{The NLO Electroweak Effects}
The dependence of the $t\bar{t}$ production cross section on the top quark Yukawa coupling arises only  when considering electroweak loop corrections. 
For the SM hypothesis, theoretical predictions for top quark pair production including electroweak corrections have been know for a long time ~\cite{Beenakker:1993yr,Kuhn:2005it,Bernreuther:2006vg,Moretti:2006nf,Kuhn:2006vh,Kuhn:2013zoa} and their implementation is available via published codes like \MCFM~\cite{Campbell:2016dks}.  Version $2.1$ of {\tt HATHOR}~\cite{Aliev:2010zk} allows for the calculation of electroweak corrections to top quark pair production with a scalable \CP-even top quark Yukawa coupling. 

In Ref.~\cite{Martini:2019lsi} the calculation of electroweak loops in hadronic $t\bar{t}$ production with modified couplings of the top quark to the electroweak gauge bosons in terms of higher-dimensional EFT operators has been presented by some of us. Building upon the techniques developed in  Ref.~\cite{Martini:2019lsi}, we allow for arbitrary \CP scenarios of the top quark Yukawa couplings parametrized by the effective Lagrangian of the interaction of the top quark $t$ and a scalar particle $H$
\begin{eqnarray} 
&& {\cal L}(Htt) = - \frac{m_t}{v} \bar{\psi}_{t} \left ( \kappa + \mathrm{i} \, \tilde\kappa \gamma_5 \right ) \psi_{t} \, H,
  \label{eq:lagrang-spin0-qq}
\end{eqnarray}
 where the $\kappa$ term is \CP even, and  the $\tilde{\kappa}$ term is \CP odd. The parameters $\kappa$ and $\tilde{\kappa}$ can be connected to the real and imaginary part of the Wilson coefficient $C^{u\varphi}_{tt}$ of the respective dimension-six operator $Q_{u\varphi}$, as defined in the Warsaw basis of the SMEFT~\cite{Dedes:2017zog}, by
\begin{eqnarray*}
\kappa&=&1-{v\over \sqrt{2}m_t}{v^2\over \Lambda^2}\textrm{Re}\left[C^{u\varphi}_{tt}\right],\\
 \tilde{\kappa}&=&-{v\over \sqrt{2}m_t}{v^2\over \Lambda^2}\textrm{Im}\left[C^{u\varphi}_{tt}\right].
\end{eqnarray*}
This effective Lagrangian incorporates additional \CP-odd states, inherent to, e.g., SUSY or two-Higgs-doublet models, while  allowing for arbitrary \CP mixing with \CP-even states, eventually recovering the SM for $\kappa=1$ and $\tilde{\kappa}=0$ (cf.  Ref.~\cite{Artoisenet:2013puc}). 
We employ the Feynman rules implied by the Lagrangian in Eq.~\ref{eq:lagrang-spin0-qq} to calculate predictions for top quark pair production including electroweak corrections while parametrizing arbitrary \CP scenarios of the top quark Yukawa coupling by $\kappa$ and $\tilde{\kappa}$.

\begin{figure}[thpb]
\begin{tabular}{c c}
\includegraphics[width=0.23\textwidth]{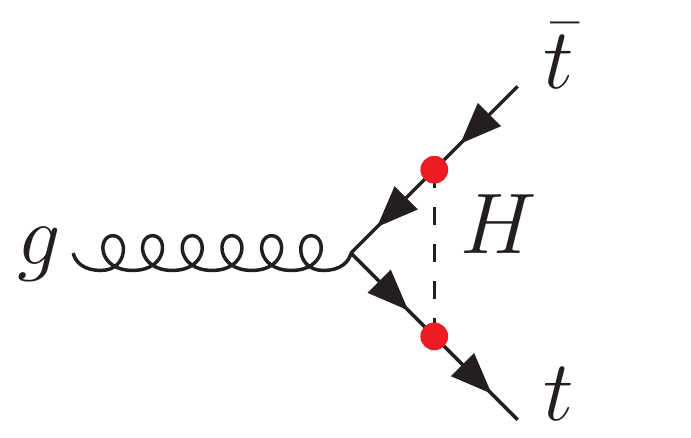} & 
\includegraphics[width=0.23\textwidth]{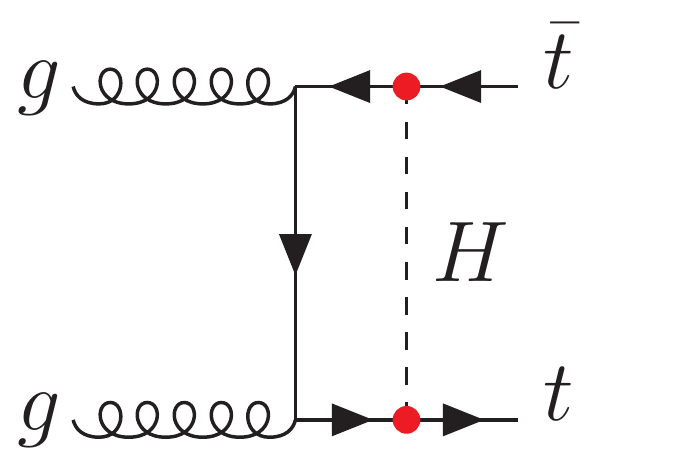} \\
(a) & (b)\\ 
\includegraphics[width=0.23\textwidth]{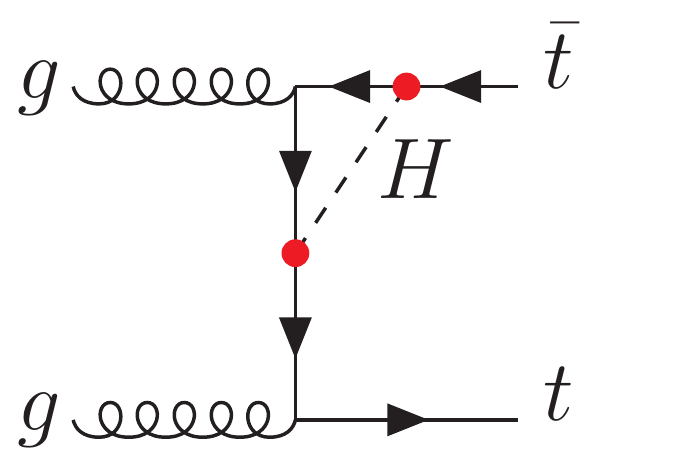} &
\includegraphics[width=0.23\textwidth]{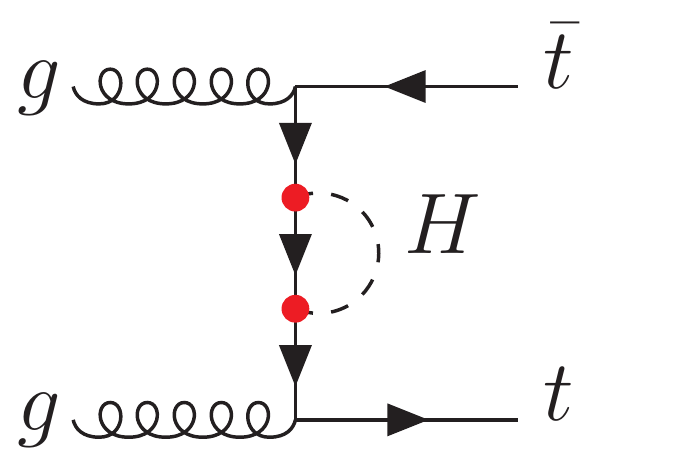} \\ 
(c) & (d)
\end{tabular}
\caption{Exemplary weak corrections from Higgs boson exchange to $\ttb$ production: Final-state vertex correction affecting the $s$-channel both in $q\bar{q}$ annihilation and gluon fusion (a). Box diagram (b), vertex correction (c) and self-energy corrections to the $t$-channel in gluon fusion.}
\label{ttbpro}
\end{figure}

\begin{figure*}[t]
\setcounter{subfigure}{0}
\centering
\subfloat{\includegraphics[width=0.38\textwidth]{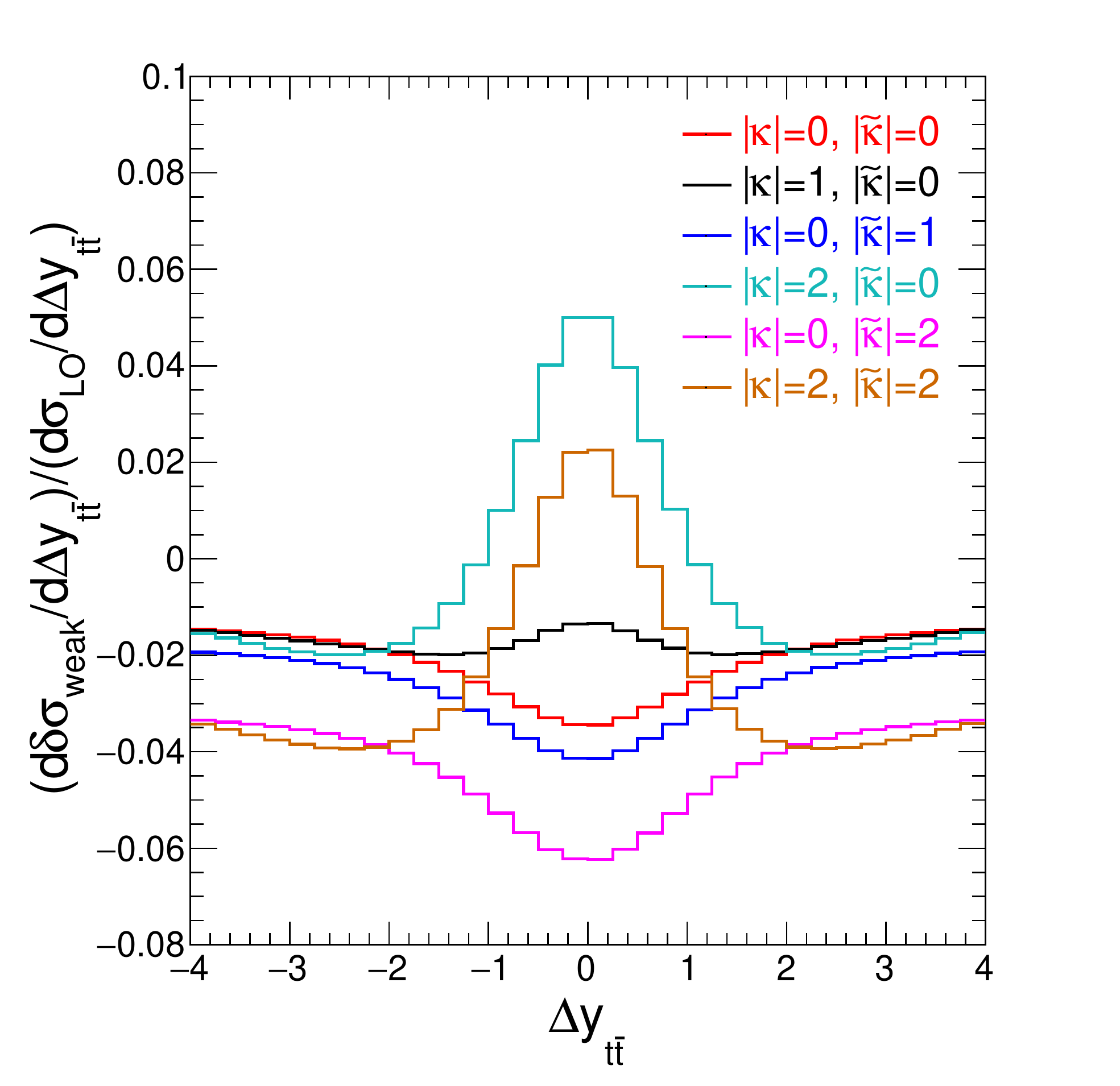}}
\subfloat{\includegraphics[width=0.38\textwidth]{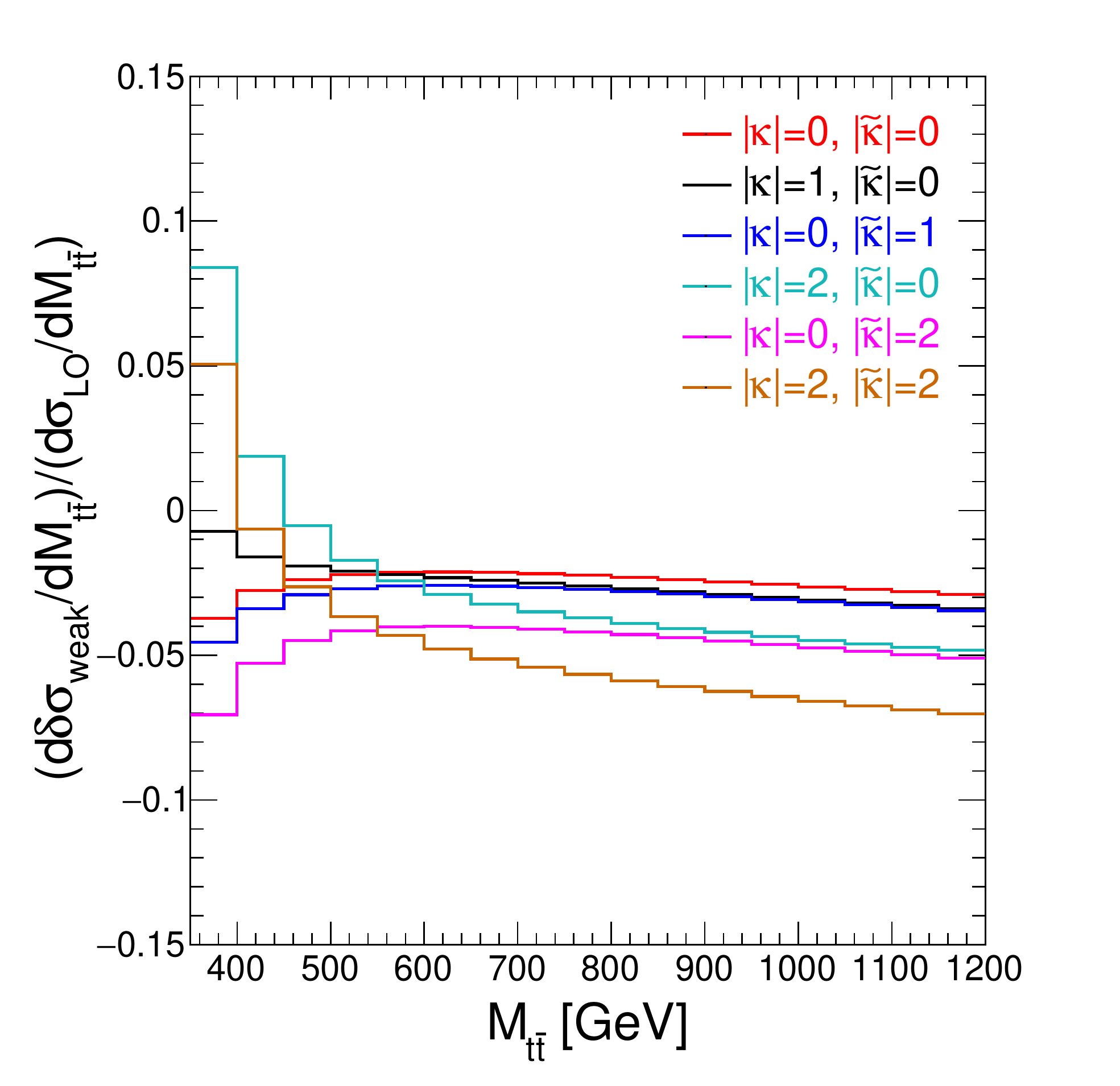}} 
\caption{The ratio of weak corrections
over the LO $\ttb$ production cross section varies with the sensitive kinematic variables $\ytt$ (left) and $\Mtt$ (right) on generator level for different anomalous parameters $\kappa$ and $\tilde{\kappa}$. The black lines correspond to the SM case.}
\label{NLO correction}
\end{figure*}

Fig.~\ref{ttbpro} shows sample diagrams for the production of a top quark pair in $q\bar{q}$ annihilation or gluon fusion including a Higgs boson running in the loop. The final state corrections shown in Fig.~\ref{ttbpro}~(a) apply to both gluonic and quark-antiquark $s$-channel production.
Due to the scalar and pseudo-scalar contributions to the top quark Yukawa coupling in Eq.~\ref{eq:lagrang-spin0-qq}, the interference terms of the tree level with the Higgs-loop diagrams are solely comprised of terms proportional to $\left(\kappa ^2+\tilde{\kappa}^2\right)$ or $\left(\kappa ^2-\tilde{\kappa}^2\right)$ when neglecting the masses of the light quarks. These terms are infrared finite but contain UV divergences. Thus, the renormalization of the top quark wave function and the top quark mass has to be consistently performed with the modified top quark Yukawa coupling in order to ensure the cancellation of the UV divergences for arbitrary values of $\kappa$ and $\tilde{\kappa}$. Following Ref.~\cite{Denner:1991kt}, we write the bare top quark field $t_0=(1+{1\over2}\delta Z_t)t$ and the bare top quark mass $m_0=m+\delta m_t$ in terms of the respective renormalized quantities $t$ and $m$ together with the renormalization constants 
\begin{eqnarray*}
\delta Z_t &=& - \mathrm{Re}\!\left[\Sigma^\mathrm{L}_t(m_t^2) +\Sigma^\mathrm{R}_t(m_t^2)\right]\nonumber\\
            && - 2 m_t^2 \frac{\partial}{\partial p^2} \mathrm{Re}\big[\Sigma^\mathrm{L}_t(p^2)+ \Sigma^\mathrm{R}_t(p^2)+ 2 \Sigma^\mathrm{S}_t(p^2) \big] \bigg|_{p^2=m_t^2}{\hspace{-0.5ex},}\hspace{5.5ex}
\\
\delta m_t &=& \frac{m_t}{2} \mathrm{Re}\!\left[ \Sigma^\mathrm{L}_t(m_t^2)+ \Sigma^\mathrm{R}_t(m_t^2) + 2 \Sigma^\mathrm{S}_t(m_t^2) \right].
\end{eqnarray*}
The terms $\Sigma^\lambda_t(p^2),\;(\lambda=\mathrm{L},\mathrm{R},\mathrm{S})$ are the chiral self energies calculated at one-loop order with the electroweak gauge bosons and the Higgs boson running in the loop. In the Higgs sector, they receive modifications with respect to the SM when  arbitrary \CP scenarios of the top quark Yukawa coupling are taken into account
\begin{eqnarray}
\label{eq:tt1}
\Sigma_{tH}^\mathrm{L} &=& {\Sigma_{tH,{\text{SM}}}^\mathrm{L}}\left(\kappa ^2+\tilde{\kappa}^2\right),\\
\Sigma_{tH}^\mathrm{R} &=& {\Sigma_{tH,{\text{SM}}}^\mathrm{R}}\left(\kappa ^2+\tilde{\kappa}^2\right),\\
\Sigma_{tH}^\mathrm{S} &=& {\Sigma_{tH,{\text{SM}}}^\mathrm{S}}\left(\kappa ^2-\tilde{\kappa}^2\right).
\label{eq:tt2}
\end{eqnarray}
After renormalization, the one-loop amplitude is UV finite for arbitrary values of $\kappa$ and $\tilde{\kappa}$ and remains without any dependence on interference terms proportional to $\kappa\tilde{\kappa}$. However, because of different contributions proportional to $(\kappa ^2+\tilde{\kappa}^2)$ as well as $(\kappa ^2-\tilde{\kappa}^2)$, the shapes of kinematic distributions are separately sensitive to $\kappa$ and $\tilde{\kappa}$.

We build upon the existing implementation of the electroweak corrections to top quark pair production in \MCFM ~\cite{Campbell:2016dks} and modify the code by the analytic results of the calculation outlined above. This extension is publicly available as an external add-on to the \MCFM~program. With the modified Monte-Carlo generator \MCFM, the relative corrections to the LO result 
\begin{equation*}
\delta_{\rm wk} = \frac{d\sigma_{\rm wk}^\mathrm{NLO} - d\sigma^\mathrm{LO}}{d\sigma^\mathrm{LO}}
\end{equation*}
can be calculated for multi-dimensional kinematic distributions dependent on $\kappa$ and $\tilde{\kappa}$.
Following Ref.~\cite{Sirunyan:2019nlw}, where the size of a pure \CP-even top Yukawa coupling was measured through the distributions of the invariant mass of the top quark pair $M_{t\bar{t}}$ and their rapidity difference, $\Delta y_{t\bar{t}}=y_t-y_{\bar{t}}$, we show the electroweak correction factor for these distributions in Fig.~\ref{NLO correction}. The rapidity difference shows a strong dependence on the \CP structure of the top quark Yukawa coupling: The \CP-even contribution increases the central region while the \CP-odd contribution decreases it. The invariant mass of the top quark pair shows dependence on the \CP structure in the threshold region as well as in the tail. Therefore, these kinematic distributions are promising candidates for probing the size of possible \CP mixtures.

\subsection{Expected sensitivity through top quark pair production}
To avoid complicated combinatorial issues, we perform the study in the semi-leptonic channel, where one top quark decays hadronically and the other decays leptonically. This final state consists of one lepton (electron or muon), missing transverse momentum, four jets from two bottom quarks and two light-flavor quarks. 
 The main background comes from single top, $V$+jets and QCD multijets processes. For simplicity, we simulate the single top process to extract the shape and rescale it to the expectation of all background processes according to the results presented in the CMS analysis~\cite{Sirunyan:2019nlw}.

In this study, the events are simulated by {\tt MadGraph5\_v2.6.4}~\cite{Alwall:2014hca} and interfaced to\\{\tt Pythia8.1}~\cite{Sjostrand:2007gs} for parton shower. The detector simulation
is implemented by\\{\tt Delphes3}~\cite{deFavereau:2013fsa} with the CMS detector setting. 
The NLO weak effects are incorporated in the generated LO events by performing a two-dimensional reweighting. The weights are obtained from the above calculation, as implemented in our publicly available extension of the \MCFM program, in the $\Mtt$-$\ytt$ phase space, and applied to the LO events based on their truth level information.
 The simulated events are normalized to the $\ttb$ production cross section of $\sigma_{\ttb}=832^{+40}_{-46}$~pb~\cite{Czakon:2017wor,Czakon:2019txp,Catani:2019hip} predicted at NNLO QCD accuracy. The cross section of single top quark production is normalized to the NLO QCD prediction ~\cite{Kidonakis:2012rm,Kant:2014oha}. 

Jets and leptons with $p_T>30$~GeV and $|\eta|<2.4$ are selected. Events are required to have four jets and exactly one lepton. Two of the four jets should be $b$-tagged jets. For the $W$ boson that decay leptonically, its transverse mass $M_\mathrm{T}(W)$ is required to be less than $140$~GeV. $M_\mathrm{T}(W)$ is defined as $M_\mathrm{T}(W) =
\sqrt{{2\pt^\ell\ptmiss[1-\cos(\Delta\phi_{\ell,\ptvecmiss})]}}$, where $\pt^\ell$ is the magnitude of the transverse momentum of the lepton and $\ptvecmiss$ denotes the missing transverse momentum. The other $W$ boson is reconstructed from the two light flavor jets.
Top quark pairs are reconstructed through a maximum likelihood method~\cite{Erdmann:2013rxa}.

\begin{figure*}[thbp]
\setcounter{subfigure}{0}
\includegraphics[width=0.40\textwidth]{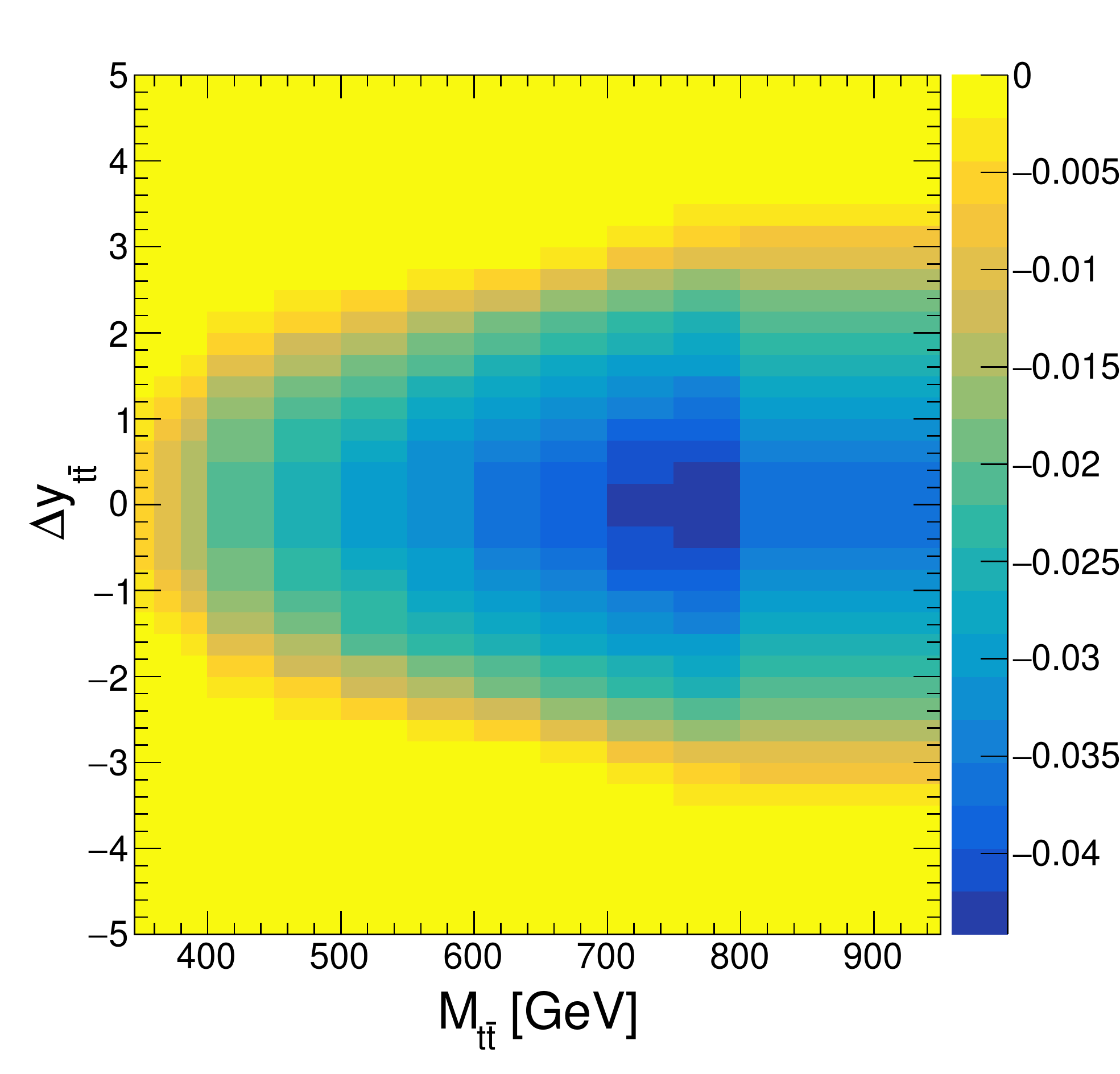}
\includegraphics[width=0.40\textwidth]{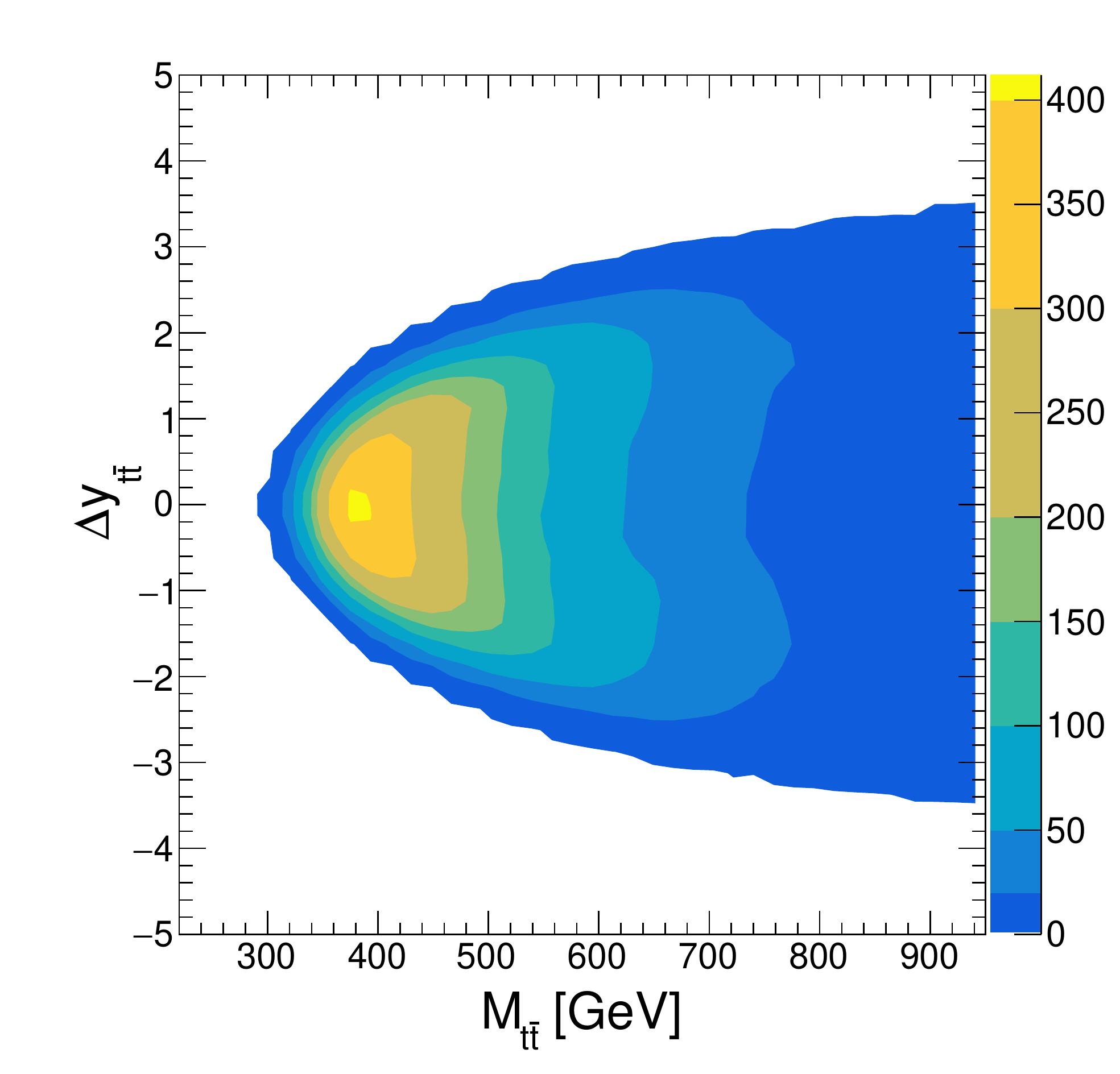} 
\caption{Left: The size of NLO weak correction over LO prediction in the two-dimensional $\Mtt$ and $\ytt$ plane, obtained at truth level with ${\kappa}=1$ and $\tilde{\kappa}=0$. 
Right: Number of simulated events in the two-dimensional plane of $\Mtt$ and $\ytt$ at reconstruction level.
}
\label{2Ddistribution}
\end{figure*}

The ratio of the weak corrections over the LO cross section in the two-dimensional plane of $\Mtt$ and $\ytt$ at parton level is shown in Fig.~\ref{2Ddistribution}. The SM couplings ($\kappa=1$, $\tilde{\kappa}=0$) are used. The corrections illustrate a correlation pattern between the two variables, thus it is important to perform the analysis in 2D. The number of events in the 2D space after event selection and reconstruction are shown in the right plot of Fig.~\ref{2Ddistribution}. 

Non-zero $\kappa$ and $\tilde{\kappa}$ values would distort the event distributions as shown in Fig~\ref{2Ddistribution}, and the effects would propagate to the reconstructed kinematics. 
The amount of \CP violation of the top quark Yukawa coupling can be quantified by the parameter 
\begin{equation*}
f_{\CP} = \frac{|\tilde\kappa|^2} {|\kappa|^2 + |\tilde\kappa|^2 }\mathrm{sign\left(\frac{\tilde{\kappa}}{\kappa}\right)},
\,
\label{eq:fa3}
\end{equation*}
which is naturally restricted to values from $-1$ to $1$. Its absolute value represents the fractional size of the \CP-odd component, and its sign reflects the relative phase of the two couplings.
We use the reconstructed 2D distribution of $\Mtt$ and $\ytt$ to extract the value of the \CP mixture parameter $f_{\CP}$. A profile likelihood method is used to obtain the expected sensitivity at 300\,$\ifb$ and 3000\,$\ifb$, respectively. Theoretical uncertainties of the signal and background processes are taken into account. In particular, QCD uncertainties of the $\ttb$ process at NNLO are about 5\% in the most relevant regions of the $\ytt$ and $\Mtt$ distributions~\cite{Czakon:2017wor}. Thus, we assign an overall 5\% uncertainty to the $\ttb$ rate. The size of the uncertainty is found to have little impact on the results. The details of the results will be discussed in Sec.~\ref{result}.

\section{On-shell sensitivity to the \CP structure of the top-Higgs coupling}
\label{thw}
\subsection{Higgs production in association with a single top quark}
The associated production of a single top quark or a top quark pair with a Higgs boson is dependent on the top quark Yukawa coupling at tree level already. Respective analyses for hadronic $\ttb H$ and $\tqH$ production are presented in Ref.~\cite{Gritsan:2016hjl}. For this work, we complete the existing results by also considering the associated production of a single top quark with a $W$ and a Higgs boson.
The $\tWH$ production at tree level has two categories of Feynman diagrams as shown in Fig.~\ref{thw process}. One category is induced by the $Htt$ coupling and the other is induced by the $HWW$ coupling. These two categories of Feynman diagrams  interfere destructively in the SM, which leads to a small total cross section of about $17$\,fb in the SM. However, \CP violation  would increase the total cross section especially when the relative sign of the $Htt$ and $HWW$ couplings flips. Single top quark production in association with a Higgs boson is also sensitive to the relative sign of the $Htt$ and $HWW$ couplings due to these interference terms. To consider arbitrary \CP scenarios for the top quark Yukawa coupling we use again the Feynman rules implied by the Lagrangian in Eq.~\ref{eq:lagrang-spin0-qq} to calculate theoretical predictions for the $\tWH$ production. 
We also include anomalous $HWW$ couplings by following the notation of Refs.~\cite{Gao:2010qx,Bolognesi:2012mm,Anderson:2013afp}
to parametrize the Lagrangian for the interaction of a scalar $H$ and two $W$ bosons
\begin{eqnarray*}
\label{eq:HVV}
 {\cal L}({H W W}) = {M^2_W  \over v} \bigg [ &&
g_{1}^{WW}  W_\mu^+ W_\mu^- 
- {g_2^{WW}\over M^2_W } W_{\mu \nu}^+  W_{\mu\nu}^-\nonumber\\
&&+{\kappa_1^{WW}\over (\Lambda_1^{WW})^2}\big( W_{\mu}^-\partial_\nu  W_{\mu\nu}^++\mathrm{h.c.}\big)\nonumber\\
 &&
- {g_4^{WW}\over M^2_W } W_{\mu \nu}^+   \tilde W_{\mu\nu}^-  
\bigg]H .
\end{eqnarray*} 
Again, the coupling parameters above have direct relations to Wilson coefficients of corresponding dimension-six operators in the Warsaw basis of the SMEFT  (cf. Refs.~\cite{deFlorian:2016spz,Gritsan:2020pib}).
Our results are incorporated in the {\tt JHUGen} thereby completing the framework's implementation of single top quark production in association with a Higgs boson with anomalous $Htt$ and $HWW$ couplings.
In this paper, the $\tWH$ production includes both the top quark associated process and the antitop quark associated process and the $HWW$ couplings are set to their SM values. 

\begin{figure}[thpb]
\setcounter{subfigure}{0}
\centering
\subfloat{\includegraphics[width=0.23\textwidth]{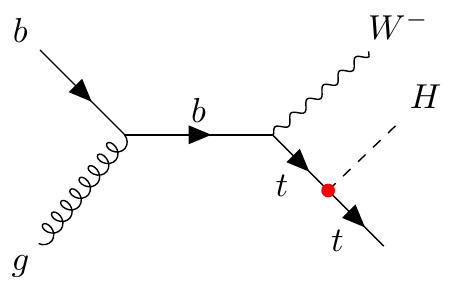}} 
\subfloat{\includegraphics[width=0.23\textwidth]{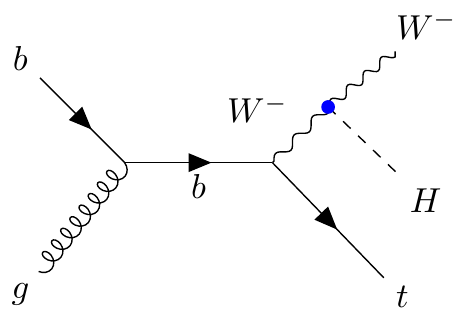}} 
\caption{Two typical Feynman diagrams of the $tHW$ production at tree level. The $Htt$ induced process (left) interferes with the $HWW$ induced process (right).}
\label{thw process}
\end{figure}

\subsection{Expected sensitivity through Higgs production in association with a single top quark }

We estimate the expected constraints on the top-Higgs \CP property in the $\tWH$ process using a matrix element method. 
The matrix element likelihood approach ({\tt MELA}) is designed to extract all essential information from the complex kinematics of a given final state. It can transform complex kinematics into a minimal set of discriminants calculated from the ratios of the matrix elements. To distinguish two different hypotheses, the ratio of probabilities ${\cal P}$ for the two hypotheses offers an optimal tool according to the Neyman-Pearson lemma~\cite{Neyman:1933wgr}. 
For measurements of properties of the Higgs boson, two types of discriminants~\cite{Anderson:2013afp,Gritsan:2016hjl} defined as below have proven to be useful
\begin{eqnarray}
\label{eq:kd0m-mela}
&& {\cal D}_{\rm alt} = \frac{{\cal P}_{\rm A}(\vec\Omega) }{{\cal P}_{\rm A}(\vec\Omega)  +{\cal P}_{B} (\vec\Omega) } \,, \\
\label{eq:kdcp-mela}
&& {\cal D}_{\rm int} = \frac{{\cal P}_{\rm int}(\vec\Omega) }{{\cal P}_{\rm A}(\vec\Omega)  +{\cal P}_{\rm B} (\vec\Omega) } \,,
\label{eq:kdbkg-mela}
\end{eqnarray}
where $\vec\Omega$ represents the 4-momenta of all particles in each final state.
The probability densities $\cal{P}$ under certain hypotheses A and B (${\cal P}_{\rm A}(\vec\Omega)$ and ${\cal P}_{\rm B}(\vec\Omega)$) 
for each event are calculated through the squared matrix element. Parton distribution functions have to be taken into account in the calculation when multiple initial parton states are concerned. $ {\cal D}_{\rm alt} $ is useful to disentangle hypotheses A and B,  while ${\cal D}_{\rm int}$, using the interference probability density ${\cal P}_{\rm int}(\vec\Omega)$ between two hypotheses, is sensitive to the interference effect. 

To estimate the sensitivity in the $pp\to \tWH$ process, we consider hadronic final states, where both the top quark and $W$ boson decay hadronically, and  $H\rightarrow\gamma\gamma$. This final state has a reasonable branching ratio and clean boson decays. The main background process is $pp\to\ttH$. Both $\ttH$ and $\tWH$ production are sensitive to the top quark Yukawa coupling, and the predicted cross sections at tree level are functions of $\kappa$ and $\tilde{\kappa}$, 

\begin{equation*}
\sigma(\kappa,\tilde{\kappa})_{\ttH}=\sigma_{\rm SM}^{\ttH}(|\kappa|^2+0.39|\tilde{\kappa}|^2),
\end{equation*} 
\begin{equation*}
\sigma(\kappa,\tilde{\kappa})_{\tWH}=\sigma_{\rm SM}^{\tWH}(2.82|\kappa|^2+2.08|\tilde{\kappa}|^2-3.87\kappa+2.05),
\end{equation*}
where $\sigma_{\rm SM}^{\ttH}$ and $\sigma_{\rm SM}^{\tWH}$ are the SM cross sections of $\ttH$ and $\tWH$ production, respectively.

We simulate $\tWH$ and $\ttH$ events using {\tt JHUGen}. 
The parton shower and hadronization are implemented by {\tt Pythia8.1} and {\tt Delphes3} is used to simulate the CMS detector response.

For event selection, we require at least $5$ jets with $p_T > 25$~GeV and $|\eta|< 2.4$, and exactly one $b$-tagged jet. Events with any isolated leptons or more than 8 jets are vetoed to remove $\ttH$ contributions. Two isolated photons are needed to pass the event selection criteria. The transverse momenta of the leading and subleading photons are required to have $p_{\rm T}^{1} > m_{\gamma\gamma}/3$ and $p_{\rm{T}}^{2}>m_{\gamma\gamma}/4$, respectively. To further suppress the $ttH$ background, we require the transverse momentum of  the Higgs boson to satisfy $p_{\mathrm{T}}^{H}> 80$~GeV. The expected number of events of the signal and background processes at 300\,$\ifb$ after selection are summarized in Table \ref{table:nevents}.
Other non-Higgs background processes like $tt+\gamma\gamma$ may be distinguished by the invariant mass of the reconstructed Higgs boson, thus are not taken into account in this study.

\begin{table}[htb]
\begin{center}
\begin{tabular}{|c|c|c|} 
 \hline
Process & Cross section [fb] & Expected number of events \\
\hline
$\tWH$ (\CP-even) & 16.8 & 0.72\\
\hline
$\tWH$ (\CP-odd) & 69.5 & 3.99\\
\hline
$\ttH$ (\CP-even) & 509.0 & 16.91 \\
 \hline
 $\ttH$ (\CP-odd)& 198.5 & 8.21 \\
 \hline
 \end{tabular}
  \caption{
Cross sections and expected number of events for signal and other contributions 
 at a luminosity of $300$~fb$^{-1}$ at $13$~ TeV. Here, \CP-even corresponds to $\kappa=1$ and $\tilde{\kappa}=0$, while \CP-odd corresponds to $\kappa=0$ and $\tilde{\kappa}=1$. The expected numbers of events are reported after event selection in the $\PH\rightarrow\gamma\gamma$ final state.
}
\label{table:nevents}
\end{center}
\end{table}

\begin{figure*}[thpb]
\setcounter{subfigure}{0}
\subfloat{\includegraphics[width=0.30\textwidth]{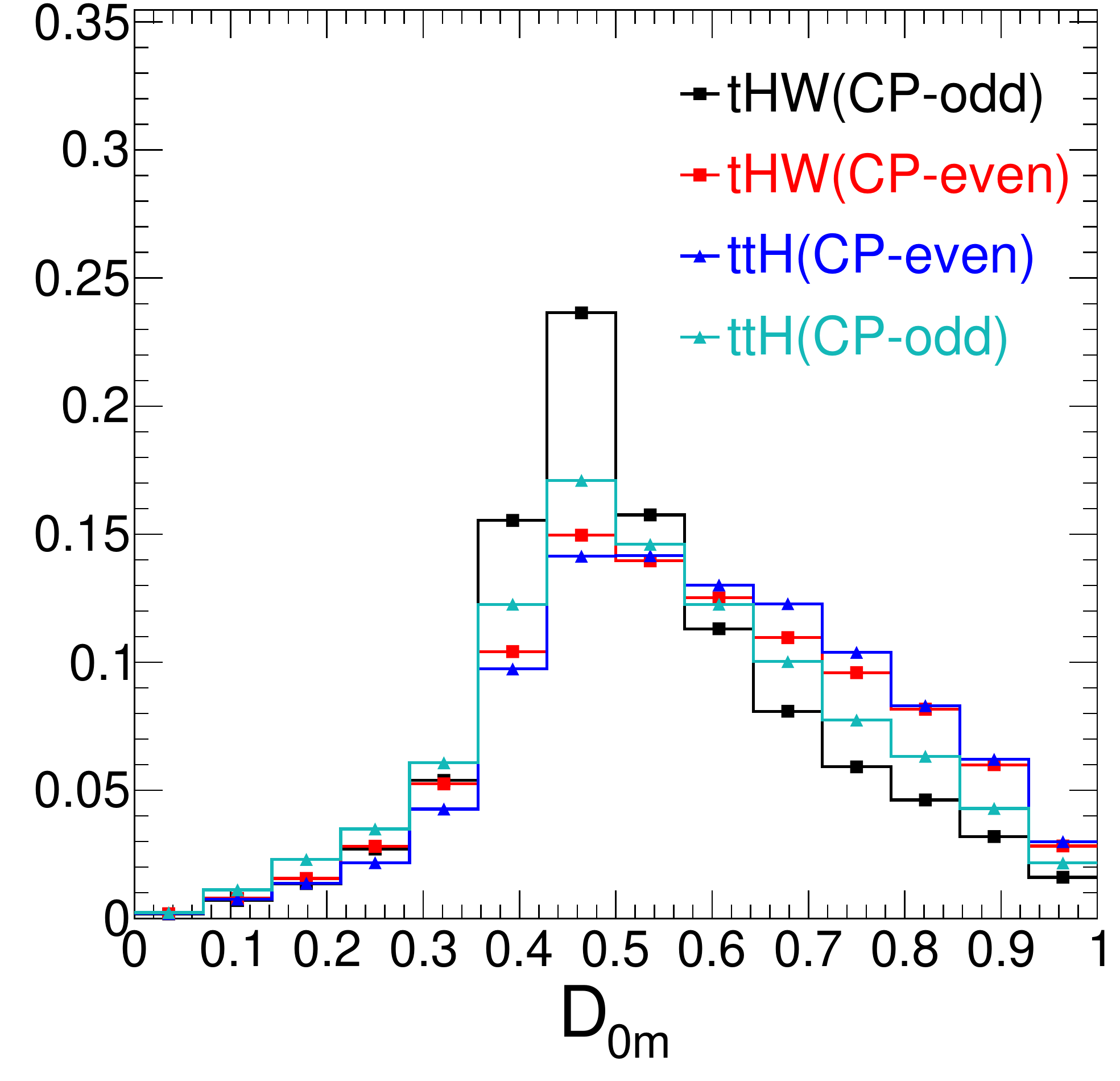}} 
\subfloat{\includegraphics[width=0.30\textwidth]{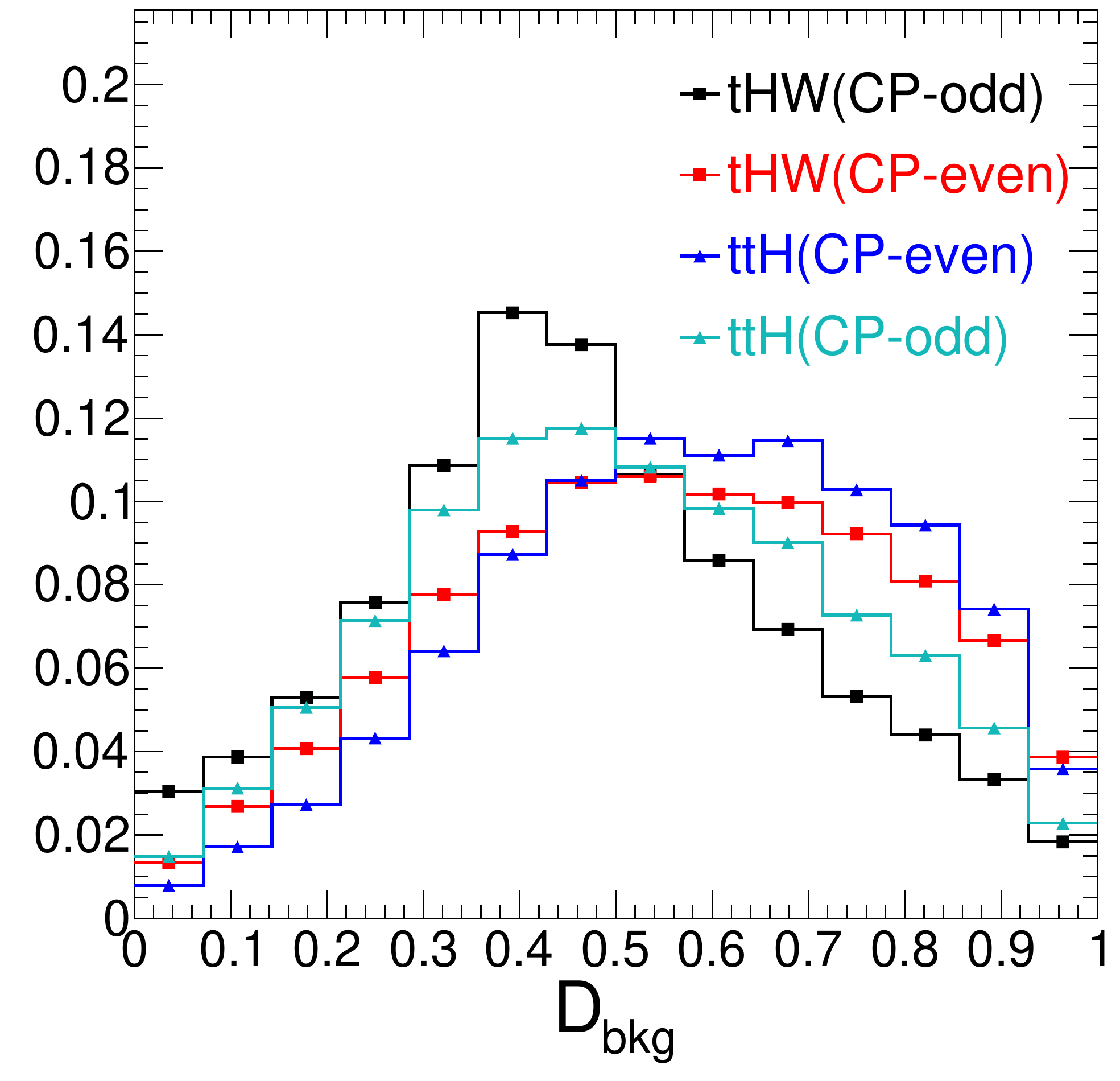}} 
\subfloat{\includegraphics[width=0.30\textwidth]{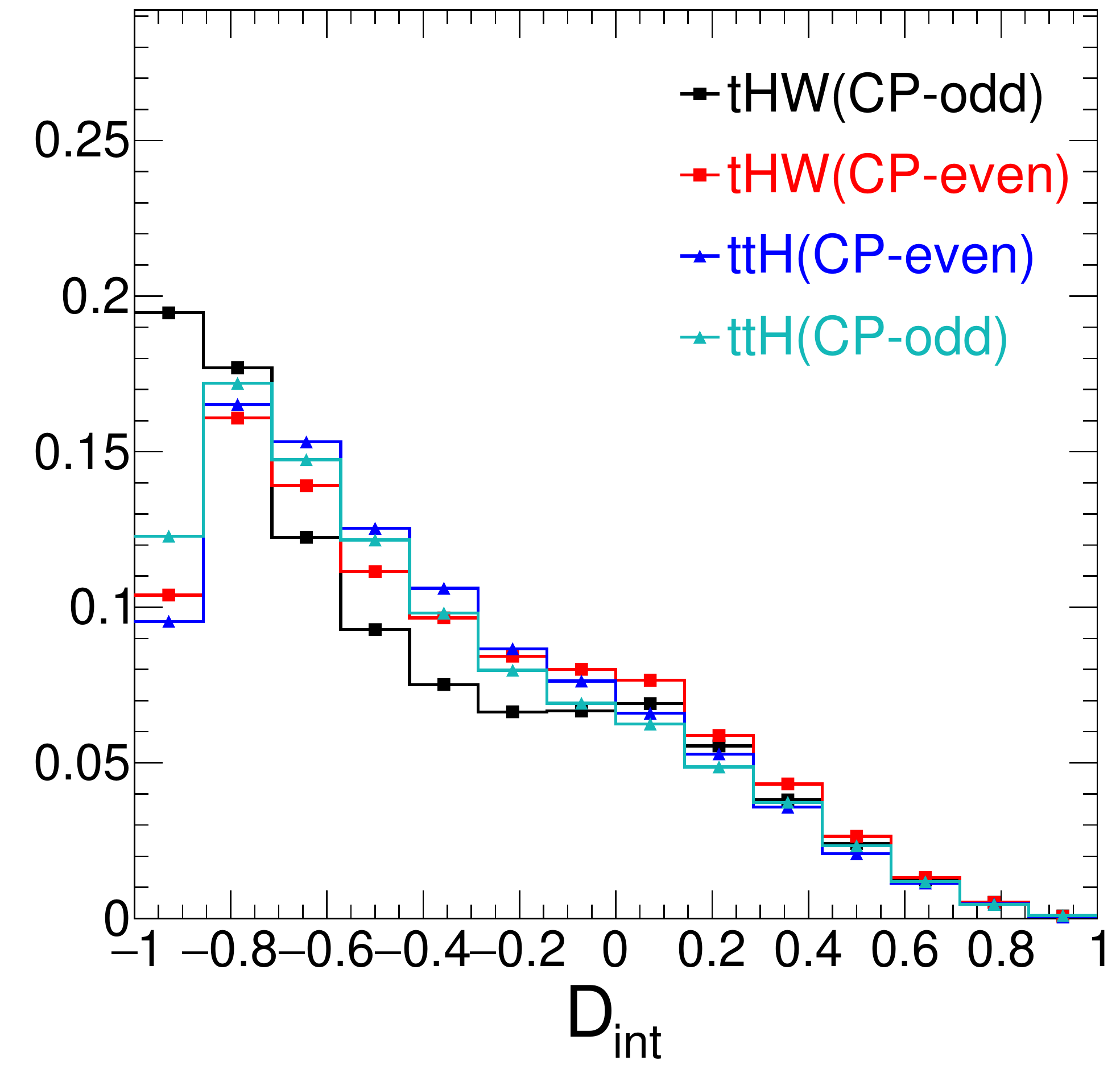}} 
\caption{The distributions of  $D_{\rm 0m}$, $D_{\rm bkg}$ and  $D_{\rm int}$. Four different scenarios are presented: the $pp\to \tWH$ process with \CP-even or \CP-odd Yukawa couplings and the $pp\to\ttH$ process with \CP-even or \CP-odd Yukawa couplings.}
\label{variables}
\end{figure*}

We apply similar techniques as used in Sec.~\ref{ttbar} to reconstruct the top quark in the final state. In total we build four matrix-element based discriminants: $\cal {D}_{\rm 0m}$, $\cal{D}_{\rm bkg}$,  $\cal D_{\CP}$ and $\cal D_{\rm int}$. The first two variables are of $\cal{D}_{\rm alt}$ type as defined in Eq.~\ref{eq:kd0m-mela}. Model A and B are \CP-even and \CP-odd model in $\cal {D}_{\rm 0m}$, and are pure $Htt$ diagram contribution and pure $HWW$ contribution in $\cal{D}_{\rm bkg}$, respectively. $\cal {D}_{\CP}$ and $D_{\rm int}$ are interference sensitive variables following Eq.~\ref{eq:kdcp-mela} definition. $\cal {D}_{\CP}$ is designed to detect the interference between \CP-even and \CP-odd $Htt$ couplings, and $\cal D_{\rm int}$ is to obtain the interference between $HWW$ and $Htt$ couplings. The distributions of $\cal{D}_{\rm 0m}$, $\cal D_{\rm bkg}$ and $\cal D_{\rm int}$ after event selection and reconstruction are presented in Fig.~\ref{variables}. The \CP-even ($\kappa=1$) and \CP-odd ($\tilde\kappa=1$) scenarios of $\tWH$ and $\ttH$ production are shown. \CP-even and \CP-odd $\tWH$ and $\ttH$ production are well separated in $D_{\rm 0m}$ and $D_{\rm bkg}$, and $D_{\rm int}$ is sensitive to the interference term of the $\tWH$ production. We use $\cal{D}_{\rm 0m}$, $\cal D_{\rm bkg}$ and $\cal D_{\rm int}$ to construct a $3$-dimensional probability density function, which is fitted to the SM distribution to estimate the \CP sensitivity.  $D_{\CP}$ is only forward-backward asymmetric in models with \CP violation as shown in shown in Fig.~\ref{dcp}, thus not used for the SM sensitivity estimation. However it will be a very powerful observable to detect any \CP violation. Unlike in the $pp\to\ttH$ process, without using the decay information of the $W$ boson and the top quark, the forward-backward asymmetry in $\cal D_{\CP}$ remains in the $\tWH$ channel. This advantage of the $pp\to \tWH$ process allows to probe the sign of the \CP violation in all top quark and $W$ decay modes.

\begin{figure}[thpb]
\setcounter{subfigure}{0}
\centering
\subfloat{\includegraphics[width=0.4\textwidth]{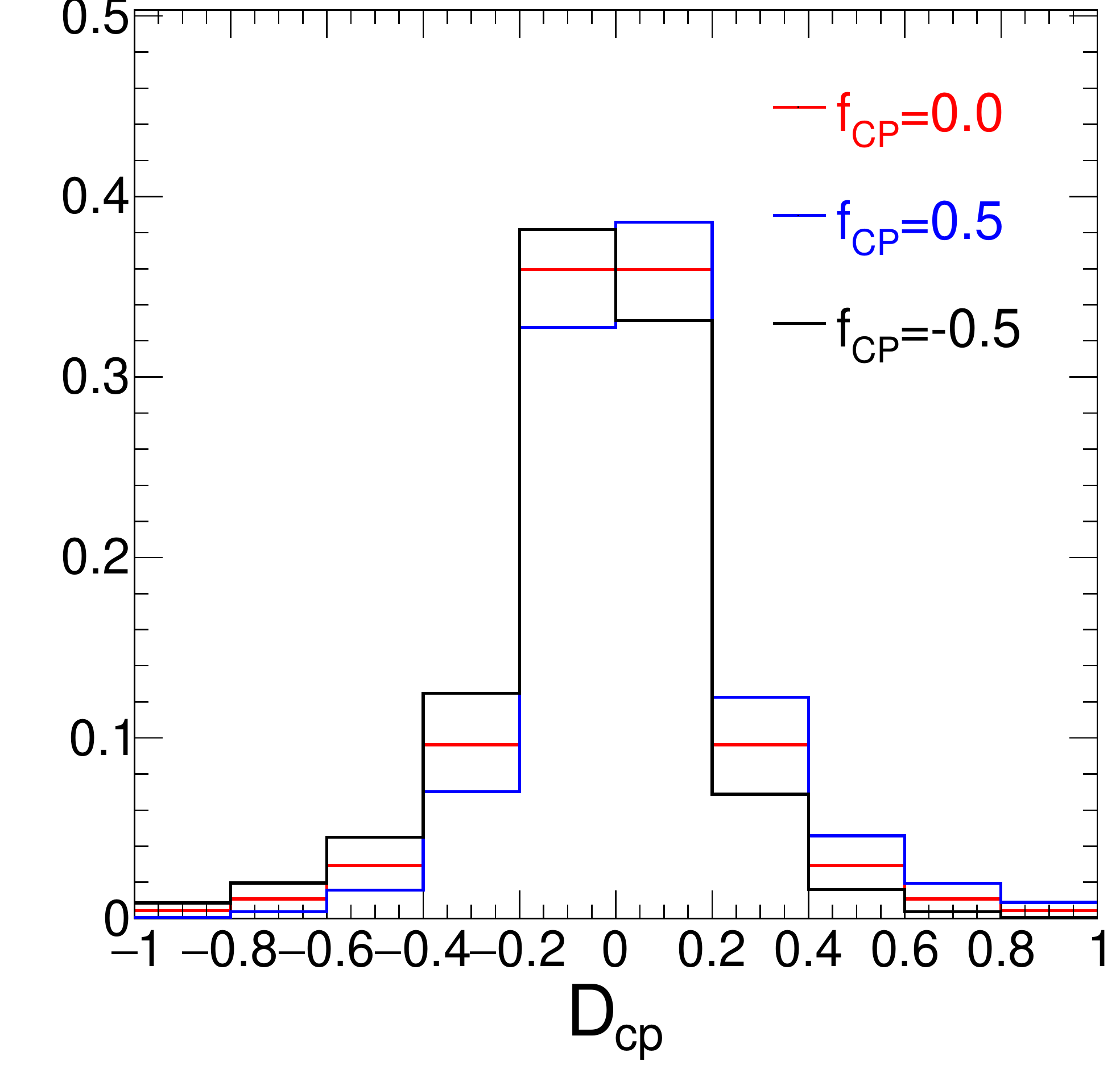}} 
\caption{The distribution of  $\cal D_{\rm CP}$ of the $pp\to \tWH$ process in three scenarios: $f_{\CP}=0$ (red), $f_{\CP}=0.5$ (blue) and $f_{\CP}=-0.5$ (black).}
\label{dcp}
\end{figure}

\section{Results}
\label{result}
A maximum likelihood fit is performed in the $\ttb$ and $\tWH$ events to quantify the sensitivity to the \CP structure of the top quark Yukawa coupling. In this study, the Higgs boson coupling to other particles except the top quark are constrained to their SM value. The $H\to \gamma\gamma$ interaction could be modified by the top quark Yukawa coupling if we assume the loop is resolved. However, only the rates of $\tWH$ and $\ttH$ will be affected. As this study focuses on the kinematic effects on the Higgs production, we assume the $H\to \gamma\gamma$ rate is the same as the SM prediction.
We present the results in two forms: one in terms of the Lagrangian coupling parameters $\kappa$ and $\tilde{\kappa}$, and the other in terms of the \CP-mixture parameter $f_{\CP}$.

The expected likelihood scan results of $\kappa$ and $\tilde{\kappa}$ at $300\,\ifb$ 
are shown in Fig.~\ref{likelihood}. The left and middle plots show the sensitivity using $\ttb$ and $\tWH$ events, respectively. The right plot shows the expected sensitivity using $tHq$ events, derived from Ref.~\cite{Gritsan:2016hjl}. It is clear from the middle and right plots that single top quark production provides sensitivity to the relative sign between the $Htt$ and $HWW$ coupling, while the $\ttb$ plot is symmetric around $\kappa=0$. This is expected as seen from Eqs.~\ref{eq:tt1}-~\ref{eq:tt2}.
\begin{figure*}[t]
\includegraphics[height=0.205\textheight,width=0.295\textwidth]{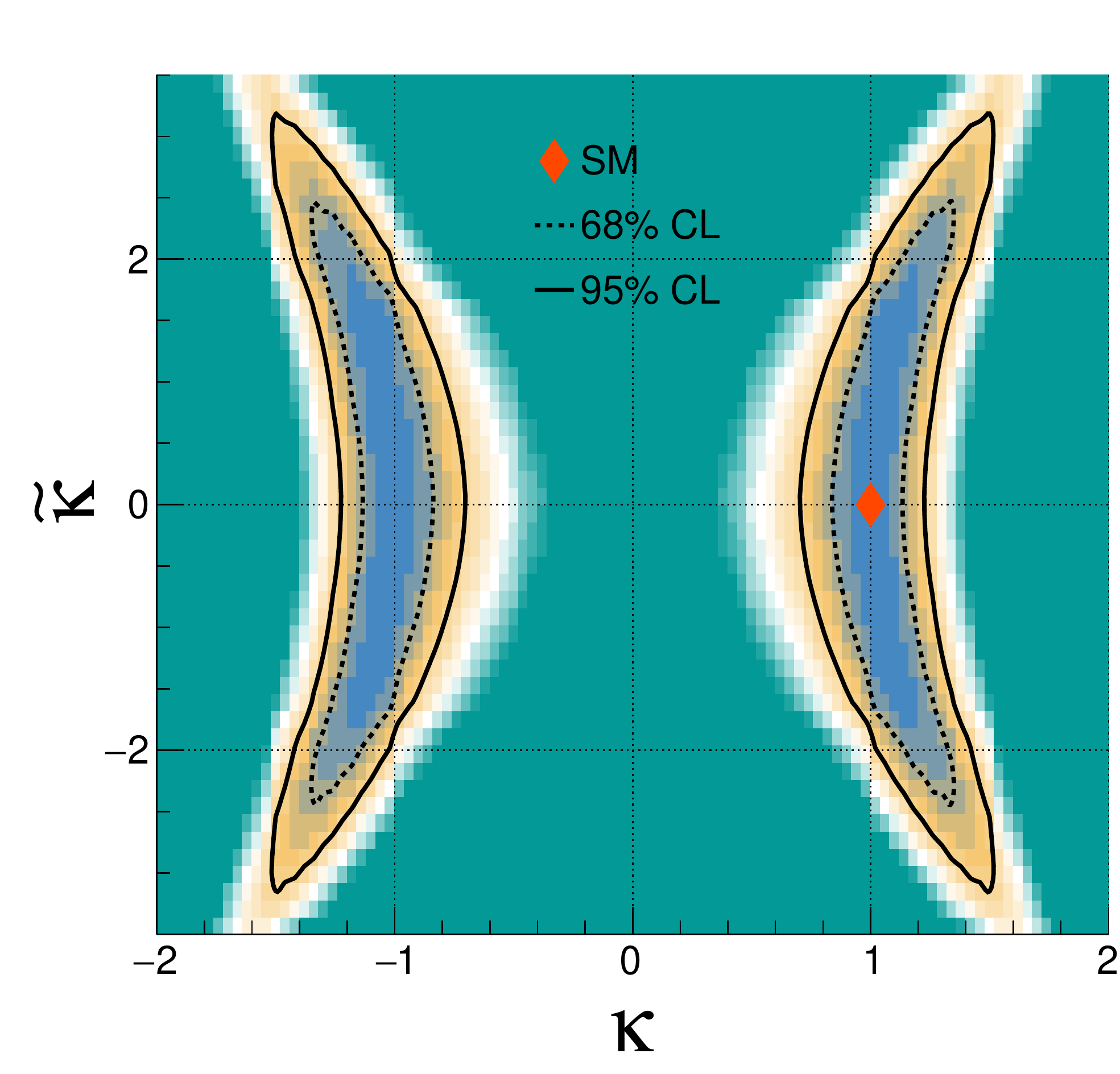}
\includegraphics[height=0.205\textheight,width=0.295\textwidth]{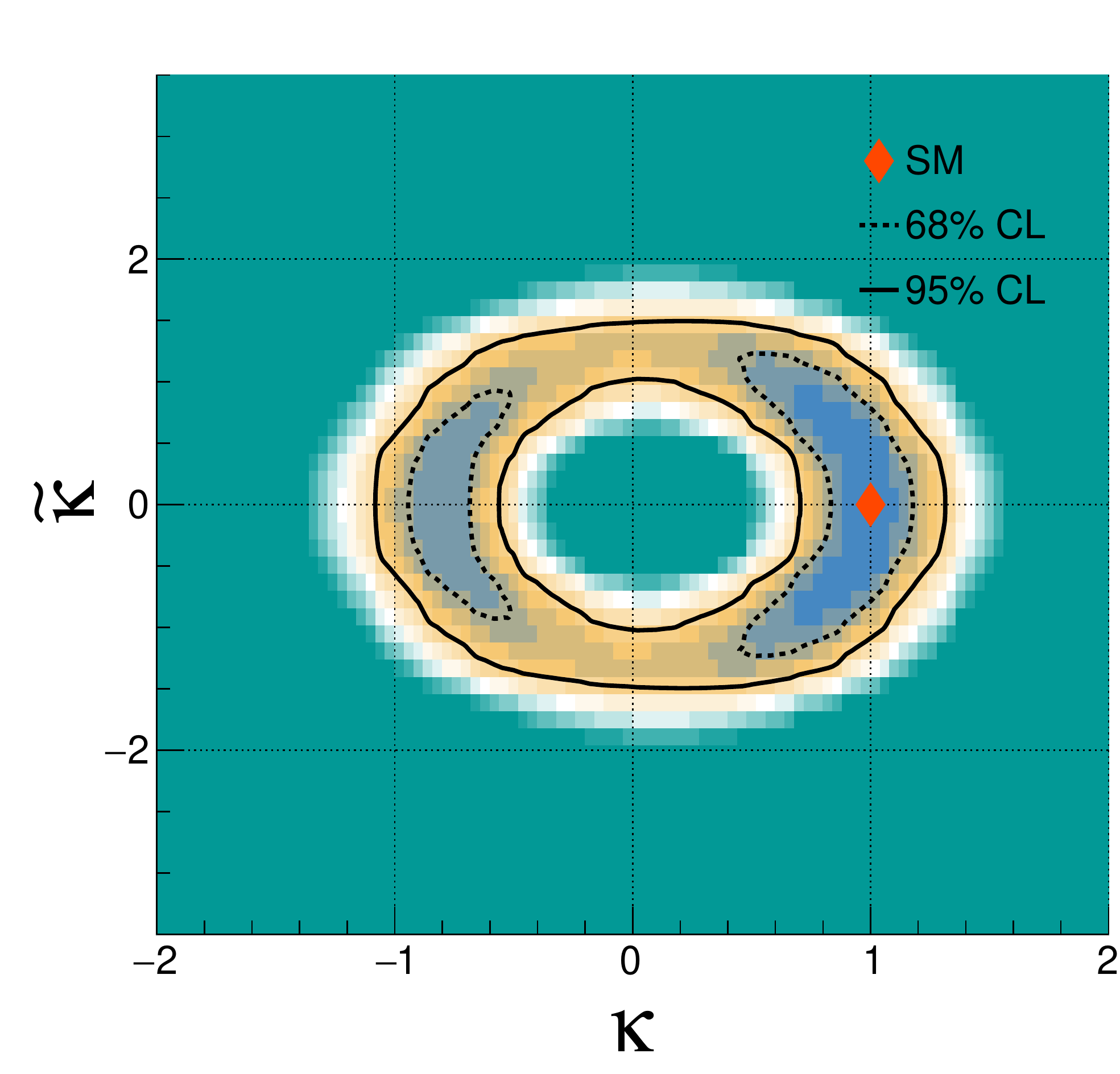}
\includegraphics[height=0.205\textheight,width=0.365\textwidth]{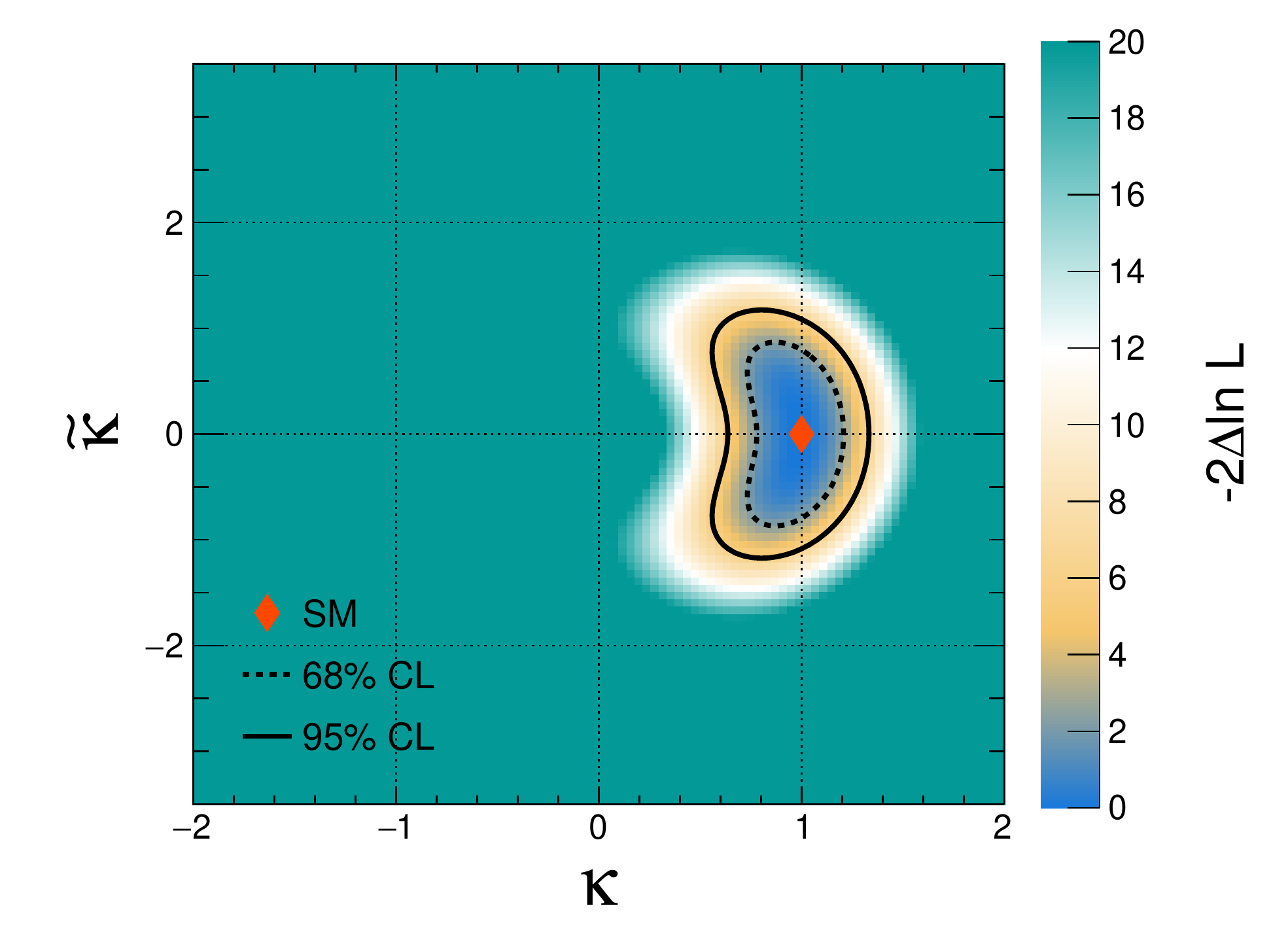}
\caption{Two dimensional likelihood scans of $\kappa$ and $\tilde{\kappa}$ in the $pp\to\ttb$ (left) and $pp\to \tWH$ (middle) and $pp\to \tqH$(right) processes at a luminosity of 300\,$\ifb$. The expected 68\% and
95\% CL regions are presented as contours with dashed and solid black lines, respectively.}
\label{likelihood}
\end{figure*}
The sensitivities of the parameter $f_{\CP}$ at the luminosity of $300\,\ifb$ and $3000\,\ifb$ are shown in Fig.~\ref{fcpscan}. When fitting the distributions in the $f_{\CP}$ framework, the overall signal rate is left unconstrained. This means whatever modification might enter $H\to \gamma\gamma$ is absorbed by this floating rate, thus the $f_{\CP}$ result is not affected by the assumptions in the $H\to \gamma\gamma$ decay. The sensitivity using $\ttH$ and $\tqH$ presented in Ref.~\cite{Gritsan:2016hjl} are also shown for comparison. One should note that while $tH$ dedicated studies aim to select $tH$ events, $\ttH$ events enter the selection due to similar final state particles. These events contribute to the \CP sensitivity in the $tH$ channel as well. Without such background, the pure contribution from the $\tWH$ events are shown as a dashed line. The curve of $\ttb$ reaches a plateau around $|{f_{\CP}}|=0.87$. This is where a switch between shape and rate effect comes into place. Beyond the boundary, the dominant effect is the overall change in the event rate with little kinematic shape variations, thus absorbed by the floating rate parameter. 

The expected sensitivities show that $\ttb$ and $tH$ events are prone to different phase spaces. They could be complementary to each other in constraining the \CP violation in top quark Yukawa coupling. Up to $300\,\ifb$, they provide rather compatible 95\% CL constraints. It will be interesting to see experimental results using all the processes.

 \begin{figure*}[thpb]
\setcounter{subfigure}{0}
\centering
\includegraphics[width=0.4\textwidth]{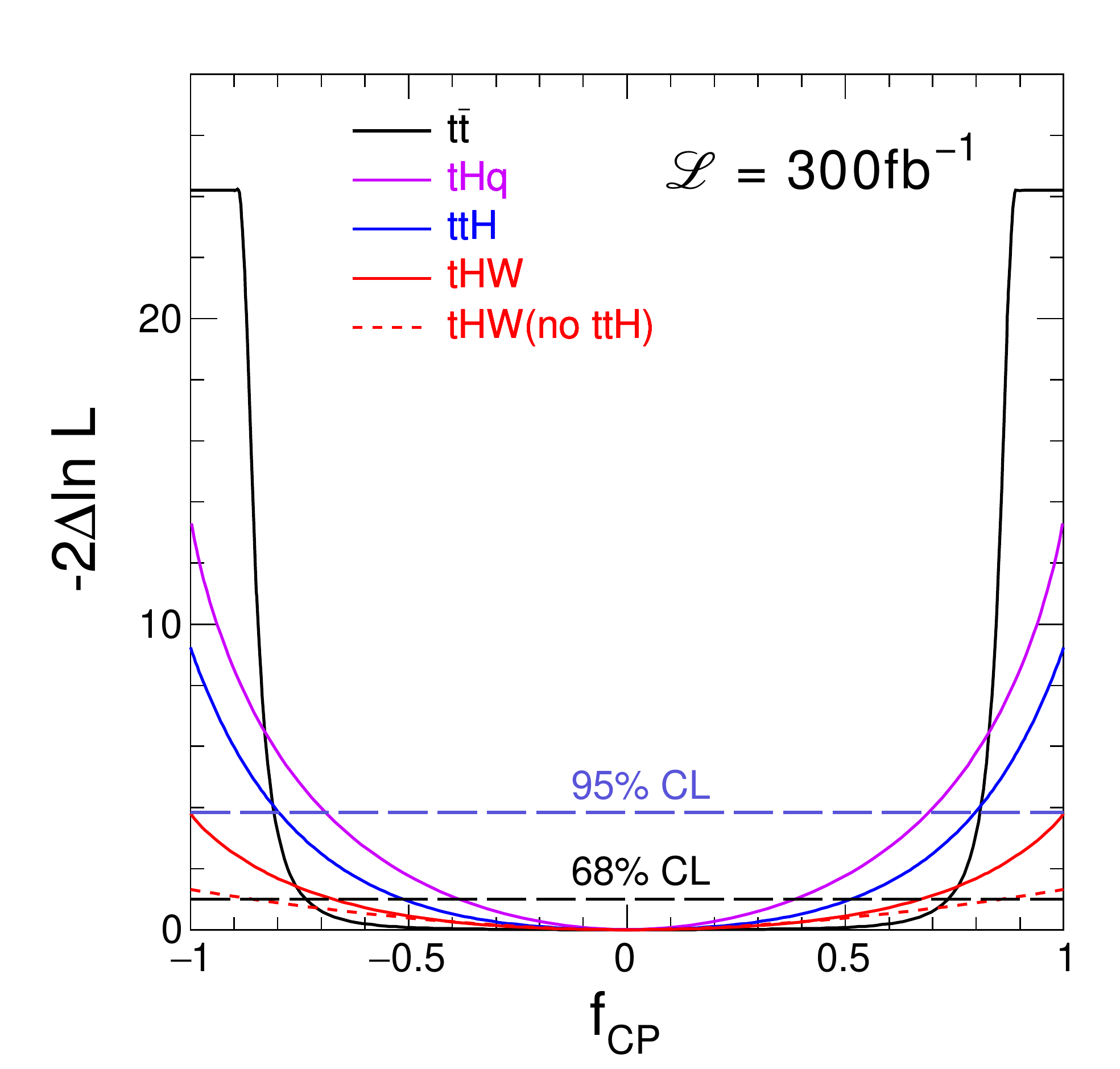}
\includegraphics[width=0.4\textwidth]{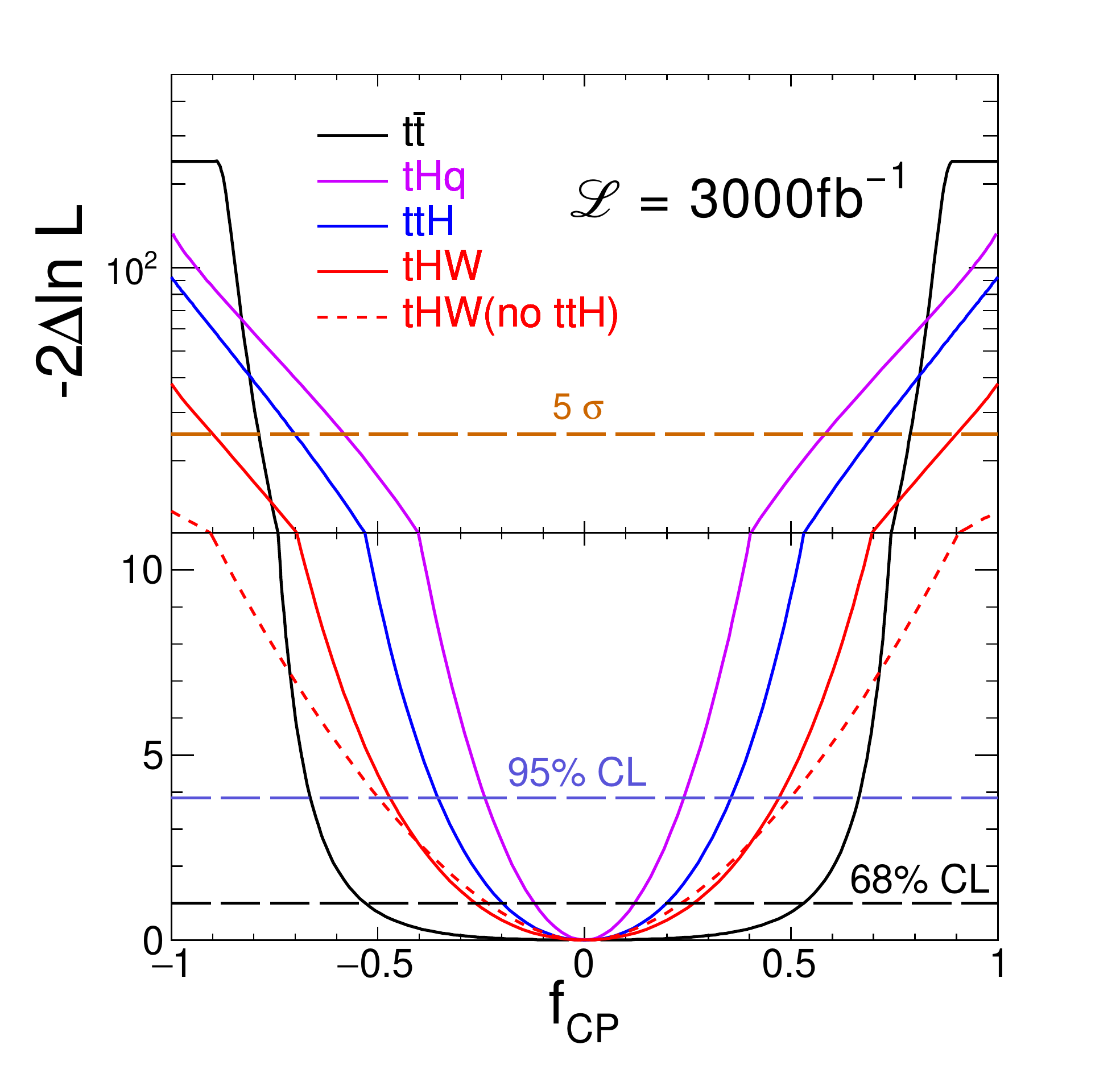}
\caption{The likelihood scan of $f_{\CP}$ at  luminosity of $300\,\ifb$ (left) and  $3000\,\ifb$ (right) with four processes shown: $pp\to\ttb$, $pp\to \tWH$, $pp\to\ttH$ and $pp\to \tqH$. The red solid line shows the expectation considering $\tWH$ and the mis-reconstructed $ttH$ events in the $\tWH$ channel, while the red dashed line assumes no contribution of $ttH$ events in the $\tWH$ channel. The black and dark blue lines represent the 68\% CL and 95\% CL lines, respectively. The results of the $\ttH$ and $\tqH$ production are cited from Ref.~\cite{Gritsan:2016hjl}. }
\label{fcpscan}
\end{figure*}
The $pp\to\ttb$ process is expected to exclude $|f_{\CP}|>0.81$ at 95\% CL at the luminosity of $300\,\ifb$. Although the total cross section of the $\tWH$ production in the SM is small, compared with the $\ttH$ and $\tqH$ production, this process can still exclude $|f_{\CP}|>0.68$ at 68\% CL and exclude the pure pseudo-scalar model at 2$\sigma$ at a luminosity of $300\,\ifb$. 
The results at a luminosity of $300\,\ifb$ can easily be projected to other luminosities such as $3000\,\ifb$ at the HL-LHC. At a luminosity of $3000\,\ifb$, the $pp\to\tWH$ process can exclude $|f_{\CP}|>0.48$ at 95\% CL, and $|f_{\CP}|>0.67$ can be excluded by the $pp\to\ttb$ process at 95\% CL. Among the four processes, the $\tqH$ production, together with the $\ttH$ events entering the selection, gives most stringent 95\% CL exclusion, which can exclude $|f_{\CP}|>0.68$ at a luminosity of $300\,\ifb$ and  $|f_{\CP}|>0.22$ at a luminosity of $3000\,\ifb$. For values of $|f_{\CP}|>0.8$, the $pp\to\ttb$ process is the best candidate for exclusion at a luminosity of $3000\,\ifb$.

\section{Summary}
\label{conc}
In this paper, we investigate the prospects of constraining the \CP structure of the coupling of the Higgs boson to the top quark through electroweak loops in $\ttb$ production. 
Sensitivity arises at loop level through off-shell degrees of freedom only. 
The fact that top quark pair production is the most dominant source of top quarks at the LHC while its theoretical description has reached impressive accuracy makes it the perfect candidate for such a novel study.
To this end, we calculate $\mathcal{O}(\alpha)$ corrections to QCD induced $\ttb$ production with arbitrary \CP-even and \CP-odd terms.
Our results for the loop sensitivity are contrasted with the ones obtained from direct on-shell probes like Higgs production in association with a single top quark or a top quark pair. 
Our results show that loop sensitivity in $\ttb$ is significantly stronger than on-shell sensitivity in associated production 
for \CP-odd admixtures $|\tilde{\kappa}| \big/ |\kappa| \ge 2.0$.
Below that value, the $pp\to\tqH$, $pp\to t\bar{t}H$ and $pp\to tHW$ processes are more sensitive.
We hope that this work demonstrates the power of loop corrections and sparks new studies. For example, in order to improve on modeling the impact of the electroweak corrections on the full kinematics in $\ttb$ production, our results could be implemented in an event generator like {\tt JHUGen}. This would allow for generation of unweighted events including electroweak  corrections together with the respective weights for different BSM hypotheses, making loop sensitivity studies possible with matrix element techniques like {\tt MELA}. 
In addition, it would be interesting to extend our calculation to $e^+e^- \to t\bar{t}$ for future collider studies.

\section*{Acknowledgments}

{We thank Andrei Gritsan for carefully reading the manuscript and his valuable feedback. We thank Mingtao Zhang for the help on validating the event generators and producing useful inputs to the  
study. We thank Klaus M\"{o}nig for valuable feedback on the implementation. This work was supported by the Fundamental Research Funds for the Central Universities (China).}

\providecommand{\href}[2]{#2}\begingroup\raggedright\endgroup


\begin{thebibliography}{10}
\providecommand{\bibinfo}[2]{#2}

\bibitem{Chatrchyan:2012ufa}
{\scshape CMS} collaboration, \emph{{Observation of a New Boson at a Mass of
  125 GeV with the CMS Experiment at the LHC}},
  \href{https://doi.org/10.1016/j.physletb.2012.08.021}{\emph{Phys. Lett. B}
  {\bfseries 716} (2012) 30} [\href{https://arxiv.org/abs/1207.7235}{{\ttfamily
  1207.7235}}].

\bibitem{Aad:2012tfa}
{\scshape ATLAS} collaboration, \emph{{Observation of a new particle in the
  search for the Standard Model Higgs boson with the ATLAS detector at the
  LHC}}, \href{https://doi.org/10.1016/j.physletb.2012.08.020}{\emph{Phys.
  Lett. B} {\bfseries 716} (2012) 1}
  [\href{https://arxiv.org/abs/1207.7214}{{\ttfamily 1207.7214}}].

\bibitem{Chatrchyan:2013lba}
{\scshape CMS} collaboration, \emph{{Observation of a New Boson with Mass Near
  125 GeV in $pp$ Collisions at $\sqrt{s}$ = 7 and 8 TeV}},
  \href{https://doi.org/10.1007/JHEP06(2013)081}{\emph{JHEP} {\bfseries 06}
  (2013) 081} [\href{https://arxiv.org/abs/1303.4571}{{\ttfamily 1303.4571}}].

\bibitem{Khachatryan:2014kca}
{\scshape CMS} collaboration, \emph{{Constraints on the spin-parity and
  anomalous HVV couplings of the Higgs boson in proton collisions at 7 and 8
  TeV}}, \href{https://doi.org/10.1103/PhysRevD.92.012004}{\emph{Phys. Rev. D}
  {\bfseries 92} (2015) 012004}
  [\href{https://arxiv.org/abs/1411.3441}{{\ttfamily 1411.3441}}].

\bibitem{Aad:2015zhl}
{\scshape ATLAS, CMS} collaboration, \emph{{Combined Measurement of the Higgs
  Boson Mass in $pp$ Collisions at $\sqrt{s}=7$ and 8 TeV with the ATLAS and
  CMS Experiments}},
  \href{https://doi.org/10.1103/PhysRevLett.114.191803}{\emph{Phys. Rev. Lett.}
  {\bfseries 114} (2015) 191803}
  [\href{https://arxiv.org/abs/1503.07589}{{\ttfamily 1503.07589}}].

\bibitem{Ginzburg:2004vp}
I.F.~Ginzburg and M.~Krawczyk, \emph{{Symmetries of two Higgs doublet model and
  CP violation}}, \href{https://doi.org/10.1103/PhysRevD.72.115013}{\emph{Phys.
  Rev. D} {\bfseries 72} (2005) 115013}
  [\href{https://arxiv.org/abs/hep-ph/0408011}{{\ttfamily hep-ph/0408011}}].

\bibitem{Gunion:2005ja}
J.F.~Gunion and H.E.~Haber, \emph{{Conditions for CP-violation in the general
  two-Higgs-doublet model}},
  \href{https://doi.org/10.1103/PhysRevD.72.095002}{\emph{Phys. Rev. D}
  {\bfseries 72} (2005) 095002}
  [\href{https://arxiv.org/abs/hep-ph/0506227}{{\ttfamily hep-ph/0506227}}].

\bibitem{Maniatis:2007vn}
M.~Maniatis, A.~von Manteuffel and O.~Nachtmann, \emph{{CP violation in the
  general two-Higgs-doublet model: A Geometric view}},
  \href{https://doi.org/10.1140/epjc/s10052-008-0712-5}{\emph{Eur. Phys. J. C}
  {\bfseries 57} (2008) 719} [\href{https://arxiv.org/abs/0707.3344}{{\ttfamily
  0707.3344}}].

\bibitem{Branco:2011iw}
G.C.~Branco, P.M.~Ferreira, L.~Lavoura, M.N.~Rebelo, M.~Sher and J.P.~Silva,
  \emph{{Theory and phenomenology of two-Higgs-doublet models}},
  \href{https://doi.org/10.1016/j.physrep.2012.02.002}{\emph{Phys. Rept.}
  {\bfseries 516} (2012) 1} [\href{https://arxiv.org/abs/1106.0034}{{\ttfamily
  1106.0034}}].

\bibitem{Haber:2012np}
H.E.~Haber and Z.~Surujon, \emph{{A Group-theoretic Condition for Spontaneous
  CP Violation}}, \href{https://doi.org/10.1103/PhysRevD.86.075007}{\emph{Phys.
  Rev. D} {\bfseries 86} (2012) 075007}
  [\href{https://arxiv.org/abs/1201.1730}{{\ttfamily 1201.1730}}].

\bibitem{Kaplan:1983sm}
D.B.~Kaplan, H.~Georgi and S.~Dimopoulos, \emph{{Composite Higgs Scalars}},
  \href{https://doi.org/10.1016/0370-2693(84)91178-X}{\emph{Phys. Lett. B}
  {\bfseries 136} (1984) 187}.

\bibitem{Georgi:1984ef}
H.~Georgi, D.B.~Kaplan and P.~Galison, \emph{{Calculation of the Composite
  Higgs Mass}}, \href{https://doi.org/10.1016/0370-2693(84)90823-2}{\emph{Phys.
  Lett. B} {\bfseries 143} (1984) 152}.

\bibitem{Georgi:1984af}
H.~Georgi and D.B.~Kaplan, \emph{{Composite Higgs and Custodial SU(2)}},
  \href{https://doi.org/10.1016/0370-2693(84)90341-1}{\emph{Phys. Lett. B}
  {\bfseries 145} (1984) 216}.

\bibitem{Dugan:1984hq}
M.J.~Dugan, H.~Georgi and D.B.~Kaplan, \emph{{Anatomy of a Composite Higgs
  Model}}, \href{https://doi.org/10.1016/0550-3213(85)90221-4}{\emph{Nucl.
  Phys. B} {\bfseries 254} (1985) 299}.

\bibitem{Contino:2003ve}
R.~Contino, Y.~Nomura and A.~Pomarol, \emph{{Higgs as a holographic
  pseudo-Goldstone boson}},
  \href{https://doi.org/10.1016/j.nuclphysb.2003.08.027}{\emph{Nucl. Phys. B}
  {\bfseries 671} (2003) 148}
  [\href{https://arxiv.org/abs/hep-ph/0306259}{{\ttfamily hep-ph/0306259}}].

\bibitem{Agashe:2004rs}
K.~Agashe, R.~Contino and A.~Pomarol, \emph{{The Minimal composite Higgs
  model}}, \href{https://doi.org/10.1016/j.nuclphysb.2005.04.035}{\emph{Nucl.
  Phys. B} {\bfseries 719} (2005) 165}
  [\href{https://arxiv.org/abs/hep-ph/0412089}{{\ttfamily hep-ph/0412089}}].

\bibitem{Giudice:2007fh}
G.F.~Giudice, C.~Grojean, A.~Pomarol and R.~Rattazzi, \emph{{The
  Strongly-Interacting Light Higgs}},
  \href{https://doi.org/10.1088/1126-6708/2007/06/045}{\emph{JHEP} {\bfseries
  06} (2007) 045} [\href{https://arxiv.org/abs/hep-ph/0703164}{{\ttfamily
  hep-ph/0703164}}].

\bibitem{Contino:2010rs}
R.~Contino, \emph{{The Higgs as a Composite Nambu-Goldstone Boson}},  in
  \emph{{Theoretical Advanced Study Institute in Elementary Particle Physics}:
  {Physics of the Large and the Small}}, pp.~235--306, 2011,
  \href{https://doi.org/10.1142/9789814327183_0005}{DOI}
  [\href{https://arxiv.org/abs/1005.4269}{{\ttfamily 1005.4269}}].

\bibitem{DeCurtis:2011yx}
S.~De~Curtis, M.~Redi and A.~Tesi, \emph{{The 4D Composite Higgs}},
  \href{https://doi.org/10.1007/JHEP04(2012)042}{\emph{JHEP} {\bfseries 04}
  (2012) 042} [\href{https://arxiv.org/abs/1110.1613}{{\ttfamily 1110.1613}}].

\bibitem{Redi:2011zi}
M.~Redi and A.~Weiler, \emph{{Flavor and CP Invariant Composite Higgs Models}},
  \href{https://doi.org/10.1007/JHEP11(2011)108}{\emph{JHEP} {\bfseries 11}
  (2011) 108} [\href{https://arxiv.org/abs/1106.6357}{{\ttfamily 1106.6357}}].

\bibitem{Mrazek:2011iu}
J.~Mrazek, A.~Pomarol, R.~Rattazzi, M.~Redi, J.~Serra and A.~Wulzer, \emph{{The
  Other Natural Two Higgs Doublet Model}},
  \href{https://doi.org/10.1016/j.nuclphysb.2011.07.008}{\emph{Nucl. Phys. B}
  {\bfseries 853} (2011) 1} [\href{https://arxiv.org/abs/1105.5403}{{\ttfamily
  1105.5403}}].

\bibitem{Redi:2012ha}
M.~Redi and A.~Tesi, \emph{{Implications of a Light Higgs in Composite
  Models}}, \href{https://doi.org/10.1007/JHEP10(2012)166}{\emph{JHEP}
  {\bfseries 10} (2012) 166} [\href{https://arxiv.org/abs/1205.0232}{{\ttfamily
  1205.0232}}].

\bibitem{Montull:2013mla}
M.~Montull, F.~Riva, E.~Salvioni and R.~Torre, \emph{{Higgs Couplings in
  Composite Models}},
  \href{https://doi.org/10.1103/PhysRevD.88.095006}{\emph{Phys. Rev. D}
  {\bfseries 88} (2013) 095006}
  [\href{https://arxiv.org/abs/1308.0559}{{\ttfamily 1308.0559}}].

\bibitem{Panico:2015jxa}
G.~Panico and A.~Wulzer, \emph{{The Composite Nambu-Goldstone Higgs}},
  vol.~913, Springer (2016),
  \href{https://doi.org/10.1007/978-3-319-22617-0}{10.1007/978-3-319-22617-0},
  [\href{https://arxiv.org/abs/1506.01961}{{\ttfamily 1506.01961}}].

\bibitem{Erdmenger:2020lvq}
J.~Erdmenger, N.~Evans, W.~Porod and K.S.~Rigatos, \emph{{Gauge/gravity
  dynamics for composite Higgs models and the top mass}},
  \href{https://doi.org/10.1103/PhysRevLett.126.071602}{\emph{Phys. Rev. Lett.}
  {\bfseries 126} (2021) 071602}
  [\href{https://arxiv.org/abs/2009.10737}{{\ttfamily 2009.10737}}].

\bibitem{Sirunyan:2020sum}
{\scshape CMS} collaboration, \emph{{Measurements of $\mathrm{t\bar{t}}H$
  Production and the CP Structure of the Yukawa Interaction between the Higgs
  Boson and Top Quark in the Diphoton Decay Channel}},
  \href{https://doi.org/10.1103/PhysRevLett.125.061801}{\emph{Phys. Rev. Lett.}
  {\bfseries 125} (2020) 061801}
  [\href{https://arxiv.org/abs/2003.10866}{{\ttfamily 2003.10866}}].

\bibitem{Aad:2020ivc}
{\scshape ATLAS} collaboration, \emph{{$CP$ Properties of Higgs Boson
  Interactions with Top Quarks in the $t\bar{t}H$ and $tH$ Processes Using $H
  \rightarrow \gamma\gamma$ with the ATLAS Detector}},
  \href{https://doi.org/10.1103/PhysRevLett.125.061802}{\emph{Phys. Rev. Lett.}
  {\bfseries 125} (2020) 061802}
  [\href{https://arxiv.org/abs/2004.04545}{{\ttfamily 2004.04545}}].

\bibitem{Czakon:2013goa}
M.~Czakon, P.~Fiedler and A.~Mitov, \emph{{Total Top-Quark Pair-Production
  Cross Section at Hadron Colliders Through $O(\alpha^4_S)$}},
  \href{https://doi.org/10.1103/PhysRevLett.110.252004}{\emph{Phys. Rev. Lett.}
  {\bfseries 110} (2013) 252004}
  [\href{https://arxiv.org/abs/1303.6254}{{\ttfamily 1303.6254}}].

\bibitem{Czakon:2015owf}
M.~Czakon, D.~Heymes and A.~Mitov, \emph{{High-precision differential
  predictions for top-quark pairs at the LHC}},
  \href{https://doi.org/10.1103/PhysRevLett.116.082003}{\emph{Phys. Rev. Lett.}
  {\bfseries 116} (2016) 082003}
  [\href{https://arxiv.org/abs/1511.00549}{{\ttfamily 1511.00549}}].

\bibitem{Czakon:2016ckf}
M.~Czakon, P.~Fiedler, D.~Heymes and A.~Mitov, \emph{{NNLO QCD predictions for
  fully-differential top-quark pair production at the Tevatron}},
  \href{https://doi.org/10.1007/JHEP05(2016)034}{\emph{JHEP} {\bfseries 05}
  (2016) 034} [\href{https://arxiv.org/abs/1601.05375}{{\ttfamily
  1601.05375}}].

\bibitem{Gao:2017goi}
J.~Gao and A.S.~Papanastasiou, \emph{{Top-quark pair-production and decay at
  high precision}},
  \href{https://doi.org/10.1103/PhysRevD.96.051501}{\emph{Phys. Rev. D}
  {\bfseries 96} (2017) 051501}
  [\href{https://arxiv.org/abs/1705.08903}{{\ttfamily 1705.08903}}].

\bibitem{Czakon:2017wor}
M.~Czakon, D.~Heymes, A.~Mitov, D.~Pagani, I.~Tsinikos and M.~Zaro,
  \emph{{Top-pair production at the LHC through NNLO QCD and NLO EW}},
  \href{https://doi.org/10.1007/JHEP10(2017)186}{\emph{JHEP} {\bfseries 10}
  (2017) 186} [\href{https://arxiv.org/abs/1705.04105}{{\ttfamily
  1705.04105}}].

\bibitem{Behring:2019iiv}
A.~Behring, M.~Czakon, A.~Mitov, A.S.~Papanastasiou and R.~Poncelet,
  \emph{{Higher order corrections to spin correlations in top quark pair
  production at the LHC}},
  \href{https://doi.org/10.1103/PhysRevLett.123.082001}{\emph{Phys. Rev. Lett.}
  {\bfseries 123} (2019) 082001}
  [\href{https://arxiv.org/abs/1901.05407}{{\ttfamily 1901.05407}}].

\bibitem{Muselli:2015kba}
C.~Muselli, M.~Bonvini, S.~Forte, S.~Marzani and G.~Ridolfi, \emph{{Top Quark
  Pair Production beyond NNLO}},
  \href{https://doi.org/10.1007/JHEP08(2015)076}{\emph{JHEP} {\bfseries 08}
  (2015) 076} [\href{https://arxiv.org/abs/1505.02006}{{\ttfamily
  1505.02006}}].

\bibitem{Piclum:2018ndt}
J.~Piclum and C.~Schwinn, \emph{{Soft-gluon and Coulomb corrections to hadronic
  top-quark pair production beyond NNLO}},
  \href{https://doi.org/10.1007/JHEP03(2018)164}{\emph{JHEP} {\bfseries 03}
  (2018) 164} [\href{https://arxiv.org/abs/1801.05788}{{\ttfamily
  1801.05788}}].

\bibitem{snowmass2021}
C.~Duhr and B.~Mistlberger, \emph{{The N3LO Frontier: Precision Predictions
  with QCDPerturbation Theory}},  2021.

\bibitem{Schmidt:1992et}
C.~R.~Schmidt and M.~E.~Peskin,
\emph{{A Probe of CP violation in top quark pair production at hadron supercolliders}},
 \href{https://journals.aps.org/prl/abstract/10.1103/PhysRevLett.69.410}{\emph{Phys. Rev. Lett.}  {\bfseries 69} (1992) 410-413}

\bibitem{Kuhn:2013zoa}
J.H.~K\"uhn, A.~Scharf and P.~Uwer, \emph{{Weak Interactions in Top-Quark Pair
  Production at Hadron Colliders: An Update}},
  \href{https://doi.org/10.1103/PhysRevD.91.014020}{\emph{Phys. Rev. D}
  {\bfseries 91} (2015) 014020}
  [\href{https://arxiv.org/abs/1305.5773}{{\ttfamily 1305.5773}}].

\bibitem{Sirunyan:2019nlw}
{\scshape CMS} collaboration, \emph{{Measurement of the top quark Yukawa
  coupling from $\mathrm{t\bar{t}}$ kinematic distributions in the lepton+jets
  final state in proton-proton collisions at $\sqrt{s} =$ 13 TeV}},
  \href{https://doi.org/10.1103/PhysRevD.100.072007}{\emph{Phys. Rev. D}
  {\bfseries 100} (2019) 072007}
  [\href{https://arxiv.org/abs/1907.01590}{{\ttfamily 1907.01590}}].

\bibitem{Gritsan:2016hjl}
A.V.~Gritsan, R.~R\"ontsch, M.~Schulze and M.~Xiao, \emph{{Constraining
  anomalous Higgs boson couplings to the heavy flavor fermions using matrix
  element techniques}},
  \href{https://doi.org/10.1103/PhysRevD.94.055023}{\emph{Phys. Rev. D}
  {\bfseries 94} (2016) 055023}
  [\href{https://arxiv.org/abs/1606.03107}{{\ttfamily 1606.03107}}].

\bibitem[{Note1()}]{Note1}
Note1, \bibinfo{note}{\protect \url {https://spin.pha.jhu.edu}.}

\bibitem[{Note2()}]{Note2}
Note2, \bibinfo{note}{\protect \url
  {https://github.com/TOPAZdevelop/MCFM-8.3\protect \_EWSMEFT\protect
  \_ADDON}}.

\bibitem{Gao:2010qx}
Y.~Gao, A.V.~Gritsan, Z.~Guo, K.~Melnikov, M.~Schulze and N.V.~Tran,
  \emph{{Spin Determination of Single-Produced Resonances at Hadron
  Colliders}}, \href{https://doi.org/10.1103/PhysRevD.81.075022}{\emph{Phys.
  Rev. D} {\bfseries 81} (2010) 075022}
  [\href{https://arxiv.org/abs/1001.3396}{{\ttfamily 1001.3396}}].

\bibitem{Bolognesi:2012mm}
S.~Bolognesi, Y.~Gao, A.V.~Gritsan, K.~Melnikov, M.~Schulze, N.V.~Tran et~al.,
  \emph{{On the spin and parity of a single-produced resonance at the LHC}},
  \href{https://doi.org/10.1103/PhysRevD.86.095031}{\emph{Phys. Rev. D}
  {\bfseries 86} (2012) 095031}
  [\href{https://arxiv.org/abs/1208.4018}{{\ttfamily 1208.4018}}].

\bibitem{Anderson:2013afp}
I.~Anderson et~al., \emph{{Constraining Anomalous HVV Interactions at Proton
  and Lepton Colliders}},
  \href{https://doi.org/10.1103/PhysRevD.89.035007}{\emph{Phys. Rev. D}
  {\bfseries 89} (2014) 035007}
  [\href{https://arxiv.org/abs/1309.4819}{{\ttfamily 1309.4819}}].

\bibitem{Gritsan:2020pib}
A.V.~Gritsan, J.~Roskes, U.~Sarica, M.~Schulze, M.~Xiao and Y.~Zhou, \emph{{New
  features in the JHU generator framework: constraining Higgs boson properties
  from on-shell and off-shell production}},
  \href{https://doi.org/10.1103/PhysRevD.102.056022}{\emph{Phys. Rev. D}
  {\bfseries 102} (2020) 056022}
  [\href{https://arxiv.org/abs/2002.09888}{{\ttfamily 2002.09888}}].

\bibitem{Khachatryan:2016tnr}
{\scshape CMS} collaboration, \emph{{Combined search for anomalous pseudoscalar
  HVV couplings in VH(H $\to b \bar b$) production and H $\to$ VV decay}},
  \href{https://doi.org/10.1016/j.physletb.2016.06.004}{\emph{Phys. Lett. B}
  {\bfseries 759} (2016) 672}
  [\href{https://arxiv.org/abs/1602.04305}{{\ttfamily 1602.04305}}].

\bibitem{Sirunyan:2017tqd}
{\scshape CMS} collaboration, \emph{{Constraints on anomalous Higgs boson
  couplings using production and decay information in the four-lepton final
  state}}, \href{https://doi.org/10.1016/j.physletb.2017.10.021}{\emph{Phys.
  Lett. B} {\bfseries 775} (2017) 1}
  [\href{https://arxiv.org/abs/1707.00541}{{\ttfamily 1707.00541}}].

\bibitem{Sirunyan:2019twz}
{\scshape CMS} collaboration, \emph{{Measurements of the Higgs boson width and
  anomalous $HVV$ couplings from on-shell and off-shell production in the
  four-lepton final state}},
  \href{https://doi.org/10.1103/PhysRevD.99.112003}{\emph{Phys. Rev. D}
  {\bfseries 99} (2019) 112003}
  [\href{https://arxiv.org/abs/1901.00174}{{\ttfamily 1901.00174}}].

\bibitem{Sirunyan:2019nbs}
{\scshape CMS} collaboration, \emph{{Constraints on anomalous $HVV$ couplings
  from the production of Higgs bosons decaying to $\tau$ lepton pairs}},
  \href{https://doi.org/10.1103/PhysRevD.100.112002}{\emph{Phys. Rev. D}
  {\bfseries 100} (2019) 112002}
  [\href{https://arxiv.org/abs/1903.06973}{{\ttfamily 1903.06973}}].

\bibitem{Campbell:2016dks}
J.M.~Campbell, D.~Wackeroth and J.~Zhou, \emph{{Study of weak corrections to
  Drell-Yan, top-quark pair, and dijet production at high energies with MCFM}},
  \href{https://doi.org/10.1103/PhysRevD.94.093009}{\emph{Phys. Rev. D}
  {\bfseries 94} (2016) 093009}
  [\href{https://arxiv.org/abs/1608.03356}{{\ttfamily 1608.03356}}].

\bibitem{Campbell:2019dru}
J.~Campbell and T.~Neumann, \emph{{Precision Phenomenology with MCFM}},
  \href{https://doi.org/10.1007/JHEP12(2019)034}{\emph{JHEP} {\bfseries 12}
  (2019) 034} [\href{https://arxiv.org/abs/1909.09117}{{\ttfamily
  1909.09117}}].

\bibitem{Bahl:2020wee}
H.~Bahl, P.~Bechtle, S.~Heinemeyer, J.~Katzy, T.~Klingl, K.~Peters et~al.,
  \emph{{Indirect $\mathcal{CP}$ probes of the Higgs-top-quark interaction:
  current LHC constraints and future opportunities}},
  \href{https://doi.org/10.1007/JHEP11(2020)127}{\emph{JHEP} {\bfseries 11}
  (2020) 127} [\href{https://arxiv.org/abs/2007.08542}{{\ttfamily
  2007.08542}}].

\bibitem{Ellis:2013yxa}
J.~Ellis, D.S.~Hwang, K.~Sakurai and M.~Takeuchi, \emph{{Disentangling
  Higgs-Top Couplings in Associated Production}},
  \href{https://doi.org/10.1007/JHEP04(2014)004}{\emph{JHEP} {\bfseries 04}
  (2014) 004} [\href{https://arxiv.org/abs/1312.5736}{{\ttfamily 1312.5736}}].

\bibitem{Demartin:2014fia}
F.~Demartin, F.~Maltoni, K.~Mawatari, B.~Page and M.~Zaro, \emph{{Higgs
  characterisation at NLO in QCD: CP properties of the top-quark Yukawa
  interaction}},
  \href{https://doi.org/10.1140/epjc/s10052-014-3065-2}{\emph{Eur. Phys. J. C}
  {\bfseries 74} (2014) 3065}
  [\href{https://arxiv.org/abs/1407.5089}{{\ttfamily 1407.5089}}].

\bibitem{Buckley:2015ctj}
M.~R.~Buckley and D.~Goncalves,
\emph{{Constraining the Strength and CP Structure of Dark Production at the LHC: the Associated Top-Pair Channel}},
\href{https://journals.aps.org/prd/abstract/10.1103/PhysRevD.93.034003}{\emph{Phys. Rev. D} {\bfseries 93} (2016) 034003}
 [\href{https://arxiv.org/abs/1511.06451}{{\ttfamily 1511.06451}}].

\bibitem{Demartin:2015uha}
F.~Demartin, F.~Maltoni, K.~Mawatari and M.~Zaro, \emph{{Higgs production in
  association with a single top quark at the LHC}},
  \href{https://doi.org/10.1140/epjc/s10052-015-3475-9}{\emph{Eur. Phys. J. C}
  {\bfseries 75} (2015) 267}
  [\href{https://arxiv.org/abs/1504.00611}{{\ttfamily 1504.00611}}].

\bibitem{Demartin:2016axk}
F.~Demartin, B.~Maier, F.~Maltoni, K.~Mawatari and M.~Zaro, \emph{{tWH
  associated production at the LHC}},
  \href{https://doi.org/10.1140/epjc/s10052-017-4601-7}{\emph{Eur. Phys. J. C}
  {\bfseries 77} (2017) 34} [\href{https://arxiv.org/abs/1607.05862}{{\ttfamily
  1607.05862}}].

\bibitem{Kobakhidze:2016mfx}
A.~Kobakhidze, N.~Liu, L.~Wu and J.~Yue, \emph{{Implications of CP-violating
  Top-Higgs Couplings at LHC and Higgs Factories}},
  \href{https://doi.org/10.1103/PhysRevD.95.015016}{\emph{Phys. Rev. D}
  {\bfseries 95} (2017) 015016}
  [\href{https://arxiv.org/abs/1610.06676}{{\ttfamily 1610.06676}}].

\bibitem{Azevedo:2017qiz}
D.~Azevedo, A.~Onofre, F.~Filthaut and R.~Gon\c{c}alo, \emph{{CP tests of Higgs
  couplings in $t\bar{t}h$ semileptonic events at the LHC}},
  \href{https://doi.org/10.1103/PhysRevD.98.033004}{\emph{Phys. Rev. D}
  {\bfseries 98} (2018) 033004}
  [\href{https://arxiv.org/abs/1711.05292}{{\ttfamily 1711.05292}}].

\bibitem{Barger:2018tqn}
V.~Barger, K.~Hagiwara and Y.-J.~Zheng, \emph{{Probing the Higgs Yukawa
  coupling to the top quark at the LHC via single top+Higgs production}},
  \href{https://doi.org/10.1103/PhysRevD.99.031701}{\emph{Phys. Rev. D}
  {\bfseries 99} (2019) 031701}
  [\href{https://arxiv.org/abs/1807.00281}{{\ttfamily 1807.00281}}].

\bibitem{Kraus:2019myc}
M.~Kraus, T.~Martini, S.~Peitzsch and P.~Uwer, \emph{{Exploring BSM Higgs
  couplings in single top-quark production}},
  \href{https://arxiv.org/abs/1908.09100}{{\ttfamily 1908.09100}}.

\bibitem{Faroughy:2019ird}
D.A.~Faroughy, J.F.~Kamenik, N.~Ko\v{s}nik and A.~Smolkovi\v{c}, \emph{{Probing
  the $CP$ nature of the top quark Yukawa at hadron colliders}},
  \href{https://doi.org/10.1007/JHEP02(2020)085}{\emph{JHEP} {\bfseries 02}
  (2020) 085} [\href{https://arxiv.org/abs/1909.00007}{{\ttfamily
  1909.00007}}].

\bibitem{Bortolato:2020zcg}
B.~Bortolato, J.F.~Kamenik, N.~Ko\v{s}nik and A.~Smolkovi\v{c},
  \emph{{Optimized probes of $CP$ -odd effects in the $t \bar t h$ process at
  hadron colliders}},
  \href{https://doi.org/10.1016/j.nuclphysb.2021.115328}{\emph{Nucl. Phys. B}
  {\bfseries 964} (2021) 115328}
  [\href{https://arxiv.org/abs/2006.13110}{{\ttfamily 2006.13110}}].

\bibitem{Brod:2013cka}
J.~Brod, U.~Haisch and J.~Zupan, \emph{{Constraints on CP-violating Higgs
  couplings to the third generation}},
  \href{https://doi.org/10.1007/JHEP11(2013)180}{\emph{JHEP} {\bfseries 11}
  (2013) 180} [\href{https://arxiv.org/abs/1310.1385}{{\ttfamily 1310.1385}}].

\bibitem{Chien:2015xha}
Y.T.~Chien, V.~Cirigliano, W.~Dekens, J.~de~Vries and E.~Mereghetti,
  \emph{{Direct and indirect constraints on CP-violating Higgs-quark and
  Higgs-gluon interactions}},
  \href{https://doi.org/10.1007/JHEP02(2016)011}{\emph{JHEP} {\bfseries 02}
  (2016) 011} [\href{https://arxiv.org/abs/1510.00725}{{\ttfamily
  1510.00725}}].

\bibitem{Cirigliano:2016nyn}
V.~Cirigliano, W.~Dekens, J.~de~Vries and E.~Mereghetti, \emph{{Constraining
  the top-Higgs sector of the Standard Model Effective Field Theory}},
  \href{https://doi.org/10.1103/PhysRevD.94.034031}{\emph{Phys. Rev. D}
  {\bfseries 94} (2016) 034031}
  [\href{https://arxiv.org/abs/1605.04311}{{\ttfamily 1605.04311}}].

\bibitem{Panico:2018hal}
G.~Panico, A.~Pomarol and M.~Riembau, \emph{{EFT approach to the electron
  Electric Dipole Moment at the two-loop level}},
  \href{https://doi.org/10.1007/JHEP04(2019)090}{\emph{JHEP} {\bfseries 04}
  (2019) 090} [\href{https://arxiv.org/abs/1810.09413}{{\ttfamily
  1810.09413}}].

\bibitem{Fuchs:2020uoc}
E.~Fuchs, M.~Losada, Y.~Nir and Y.~Viernik, \emph{{$CP$ violation from $\tau$,
  $t$ and $b$ dimension-6 Yukawa couplings - interplay of baryogenesis, EDM and
  Higgs physics}}, \href{https://doi.org/10.1007/JHEP05(2020)056}{\emph{JHEP}
  {\bfseries 05} (2020) 056}
  [\href{https://arxiv.org/abs/2003.00099}{{\ttfamily 2003.00099}}].

\bibitem{Beenakker:1993yr}
W.~Beenakker, A.~Denner, W.~Hollik, R.~Mertig, T.~Sack and D.~Wackeroth,
  \emph{{Electroweak one loop contributions to top pair production in hadron
  colliders}}, \href{https://doi.org/10.1016/0550-3213(94)90454-5}{\emph{Nucl.
  Phys. B} {\bfseries 411} (1994) 343}.

\bibitem{Kuhn:2005it}
J.H.~K\"uhn, A.~Scharf and P.~Uwer, \emph{{Electroweak corrections to top-quark
  pair production in quark-antiquark annihilation}},
  \href{https://doi.org/10.1140/epjc/s2005-02423-6}{\emph{Eur. Phys. J. C}
  {\bfseries 45} (2006) 139}
  [\href{https://arxiv.org/abs/hep-ph/0508092}{{\ttfamily hep-ph/0508092}}].

\bibitem{Bernreuther:2006vg}
W.~Bernreuther, M.~Fuecker and Z.-G.~Si, \emph{{Weak interaction corrections to
  hadronic top quark pair production}},
  \href{https://doi.org/10.1103/PhysRevD.74.113005}{\emph{Phys. Rev. D}
  {\bfseries 74} (2006) 113005}
  [\href{https://arxiv.org/abs/hep-ph/0610334}{{\ttfamily hep-ph/0610334}}].

\bibitem{Moretti:2006nf}
S.~Moretti, M.~Nolten and D.~Ross, \emph{{Weak corrections to gluon-induced
  top-antitop hadro-production}},
  \href{https://doi.org/10.1016/j.physletb.2006.06.078}{\emph{Phys. Lett. B}
  {\bfseries 639} (2006) 513}
  [\href{https://arxiv.org/abs/hep-ph/0603083}{{\ttfamily hep-ph/0603083}}].

\bibitem{Kuhn:2006vh}
J.H.~K\"uhn, A.~Scharf and P.~Uwer, \emph{{Electroweak effects in top-quark
  pair production at hadron colliders}},
  \href{https://doi.org/10.1140/epjc/s10052-007-0275-x}{\emph{Eur. Phys. J. C}
  {\bfseries 51} (2007) 37}
  [\href{https://arxiv.org/abs/hep-ph/0610335}{{\ttfamily hep-ph/0610335}}].

\bibitem{Aliev:2010zk}
M.~Aliev, H.~Lacker, U.~Langenfeld, S.~Moch, P.~Uwer and M.~Wiedermann,
  \emph{{HATHOR: HAdronic Top and Heavy quarks crOss section calculatoR}},
  \href{https://doi.org/10.1016/j.cpc.2010.12.040}{\emph{Comput. Phys. Commun.}
  {\bfseries 182} (2011) 1034}
  [\href{https://arxiv.org/abs/1007.1327}{{\ttfamily 1007.1327}}].

\bibitem{Martini:2019lsi}
T.~Martini and M.~Schulze, \emph{{Electroweak loops as a probe of new physics
  in $ t\overline{t} $ production at the LHC}},
  \href{https://doi.org/10.1007/JHEP04(2020)017}{\emph{JHEP} {\bfseries 04}
  (2020) 017} [\href{https://arxiv.org/abs/1911.11244}{{\ttfamily
  1911.11244}}].

\bibitem{Dedes:2017zog}
A.~Dedes, W.~Materkowska, M.~Paraskevas, J.~Rosiek and K.~Suxho, \emph{{Feynman
  rules for the Standard Model Effective Field Theory in R$_{\xi}$ -gauges}},
  \href{https://doi.org/10.1007/JHEP06(2017)143}{\emph{JHEP} {\bfseries 06}
  (2017) 143} [\href{https://arxiv.org/abs/1704.03888}{{\ttfamily
  1704.03888}}].

\bibitem{Artoisenet:2013puc}
P.~Artoisenet et~al., \emph{{A framework for Higgs characterisation}},
  \href{https://doi.org/10.1007/JHEP11(2013)043}{\emph{JHEP} {\bfseries 11}
  (2013) 043} [\href{https://arxiv.org/abs/1306.6464}{{\ttfamily 1306.6464}}].

\bibitem{Denner:1991kt}
A.~Denner, \emph{{Techniques for calculation of electroweak radiative
  corrections at the one loop level and results for W physics at LEP-200}},
  \href{https://doi.org/10.1002/prop.2190410402}{\emph{Fortsch. Phys.}
  {\bfseries 41} (1993) 307} [\href{https://arxiv.org/abs/0709.1075}{{\ttfamily
  0709.1075}}].

\bibitem{Alwall:2014hca}
J.~Alwall, R.~Frederix, S.~Frixione, V.~Hirschi, F.~Maltoni, O.~Mattelaer
  et~al., \emph{{The automated computation of tree-level and next-to-leading
  order differential cross sections, and their matching to parton shower
  simulations}}, \href{https://doi.org/10.1007/JHEP07(2014)079}{\emph{JHEP}
  {\bfseries 07} (2014) 079} [\href{https://arxiv.org/abs/1405.0301}{{\ttfamily
  1405.0301}}].

\bibitem{Sjostrand:2007gs}
T.~Sjostrand, S.~Mrenna and P.Z.~Skands, \emph{{A Brief Introduction to PYTHIA
  8.1}}, \href{https://doi.org/10.1016/j.cpc.2008.01.036}{\emph{Comput. Phys.
  Commun.} {\bfseries 178} (2008) 852}
  [\href{https://arxiv.org/abs/0710.3820}{{\ttfamily 0710.3820}}].

\bibitem{deFavereau:2013fsa}
{\scshape DELPHES 3} collaboration, \emph{{DELPHES 3, A modular framework for
  fast simulation of a generic collider experiment}},
  \href{https://doi.org/10.1007/JHEP02(2014)057}{\emph{JHEP} {\bfseries 02}
  (2014) 057} [\href{https://arxiv.org/abs/1307.6346}{{\ttfamily 1307.6346}}].

\bibitem{Czakon:2019txp}
M.L.~Czakon et~al., \emph{{Top quark pair production at complete NLO accuracy
  with NNLO+NNLL' corrections in QCD}},
  \href{https://doi.org/10.1088/1674-1137/44/8/083104}{\emph{Chin. Phys. C}
  {\bfseries 44} (2020) 083104}
  [\href{https://arxiv.org/abs/1901.08281}{{\ttfamily 1901.08281}}].

\bibitem{Catani:2019hip}
S.~Catani, S.~Devoto, M.~Grazzini, S.~Kallweit and J.~Mazzitelli,
  \emph{{Top-quark pair production at the LHC: Fully differential QCD
  predictions at NNLO}},
  \href{https://doi.org/10.1007/JHEP07(2019)100}{\emph{JHEP} {\bfseries 07}
  (2019) 100} [\href{https://arxiv.org/abs/1906.06535}{{\ttfamily
  1906.06535}}].

\bibitem{Kidonakis:2012rm}
N.~Kidonakis, \emph{{NNLL threshold resummation for top-pair and single-top
  production}}, \href{https://doi.org/10.1134/S1063779614040091}{\emph{Phys.
  Part. Nucl.} {\bfseries 45} (2014) 714}
  [\href{https://arxiv.org/abs/1210.7813}{{\ttfamily 1210.7813}}].

\bibitem{Kant:2014oha}
P.~Kant, O.~Kind, T.~Kintscher, T.~Lohse, T.~Martini, S.~M\"olbitz et~al.,
  \emph{{HatHor for single top-quark production: Updated predictions and
  uncertainty estimates for single top-quark production in hadronic
  collisions}}, \href{https://doi.org/10.1016/j.cpc.2015.02.001}{\emph{Comput.
  Phys. Commun.} {\bfseries 191} (2015) 74}
  [\href{https://arxiv.org/abs/1406.4403}{{\ttfamily 1406.4403}}].

\bibitem{Erdmann:2013rxa}
J.~Erdmann, S.~Guindon, K.~Kroeninger, B.~Lemmer, O.~Nackenhorst, A.~Quadt
  et~al., \emph{{A likelihood-based reconstruction algorithm for top-quark
  pairs and the KLFitter framework}},
  \href{https://doi.org/10.1016/j.nima.2014.02.029}{\emph{Nucl. Instrum. Meth.
  A} {\bfseries 748} (2014) 18}
  [\href{https://arxiv.org/abs/1312.5595}{{\ttfamily 1312.5595}}].

\bibitem{deFlorian:2016spz}
{\scshape LHC Higgs Cross Section Working Group} collaboration, \emph{{Handbook
  of LHC Higgs Cross Sections: 4. Deciphering the Nature of the Higgs Sector}},
   \href{https://arxiv.org/abs/1610.07922}{{\ttfamily 1610.07922}}.

\bibitem{Neyman:1933wgr}
J.~Neyman and E.S.~Pearson, \emph{{On the Problem of the Most Efficient Tests
  of Statistical Hypotheses}},
  \href{https://doi.org/10.1098/rsta.1933.0009}{\emph{Phil. Trans. Roy. Soc.
  Lond. A} {\bfseries 231} (1933) 289}.

\end{thebibliography}
\end{document}